\documentclass[useAMS,usenatbib]{mnras}
\usepackage{graphicx,amsmath}
\usepackage{amssymb}
\usepackage{natbib}
\usepackage{hyperref}
\def\CIVdblt{{\rm C~}\kern 0.1em{\sc iv}~$\lambda\lambda 1548, 1550$}
\def\MgIIdblt{{\rm Mg~}\kern 0.1em{\sc ii}~$\lambda\lambda 2796, 2803$}
\def\NVdblt{{\rm N~}\kern 0.1em{\sc v}~$\lambda\lambda 1238, 1242$}
\def\OVIdblt{{\rm O~}\kern 0.1em{\sc vi}~$ 1031, 1037$}
\def\SiIVdblt{{\rm Si~}\kern 0.1em{\sc iv}~$\lambda\lambda1394, 1403$}
\def\AlIIIdblt{{\rm Al~}\kern 0.1em{\sc iii}~$\lambda\lambda1855,1863$}
\def\FeIIdblt{{\rm Fe~}\kern 0.1em{\sc ii}~$\lambda\lambda 2383, 2600$}
\def\NeVIIIdblt{{\rm Ne~}\kern 0.1em{\sc viii}~$ 770, 780$}

\def\NeVIII{\hbox{{\rm Ne~}\kern 0.1em{\sc viii}}}
\def\OI{\hbox{{\rm O~}\kern 0.1em{\sc i}}}
\def\OII{\hbox{{\rm O~}\kern 0.1em{\sc ii}}}
\def\OIII{\hbox{{\rm O~}\kern 0.1em{\sc iii}}}
\def\OIV{\hbox{{\rm O~}\kern 0.1em{\sc iv}}}
\def\OV{\hbox{{\rm O~}\kern 0.1em{\sc v}}}
\def\OVI{\hbox{{\rm O~}\kern 0.1em{\sc vi}}}
\def\OVII{\hbox{{\rm O~}\kern 0.1em{\sc vii}}}
\def\OVIII{\hbox{{\rm O~}\kern 0.1em{\sc viii}}}
\def\NIII{\hbox{{\rm N~}\kern 0.1em{\sc iii}}}
\def\NIV{\hbox{{\rm N~}\kern 0.1em{\sc iv}}}
\def\NVII{\hbox{{\rm N~}\kern 0.1em{\sc vii}}}
\def\CIII{\hbox{{\rm C~}\kern 0.1em{\sc iii}}}
\def\SiIII{\hbox{{\rm Si~}\kern 0.1em{\sc iii}}}
\def\SVI{\hbox{{\rm S~}\kern 0.1em{\sc vi}}}
\def\SV{\hbox{{\rm S~}\kern 0.1em{\sc v}}}
\def\NeIX{\hbox{{\rm Ne~}\kern 0.1em{\sc ix}}}

\def\AlII{\hbox{{\rm Al~}\kern 0.1em{\sc ii}}}
\def\AlIII{\hbox{{\rm Al~}\kern 0.1em{\sc iii}}}
\def\CaI{\hbox{{\rm Ca}\kern 0.1em{\sc i}}}
\def\CaII{\hbox{{\rm Ca}\kern 0.1em{\sc ii}}}
\def\CrII{\hbox{{\rm Cr}\kern 0.1em{\sc ii}}}
\def\CII{\hbox{{\rm C~}\kern 0.1em{\sc ii}}}
\def\CIII{\hbox{{\rm C~}\kern 0.1em{\sc iii}}}
\def\CIV{\hbox{{\rm C~}\kern 0.1em{\sc iv}}}
\def\CV{\hbox{{\rm C~}\kern 0.1em{\sc v}}}
\def\H{\hbox{{\rm \small{H}}}}
\def\HI{\hbox{{\rm H~}\kern 0.1em{\sc i}}}
\def\HII{\hbox{{\rm H~}\kern 0.1em{\sc ii}}}
\def\Lya{\hbox{{\rm Ly}\kern 0.1em$\alpha$}}
\def\Lyb{\hbox{{\rm Ly}\kern 0.1em$\beta$}}
\def\Lyg{\hbox{{\rm Ly}\kern 0.1em$\gamma$}}
\def\Lyd{\hbox{{\rm Ly}\kern 0.1em$\delta$}}
\def\Lyth{\hbox{{\rm Ly}\kern 0.1em$\theta$}}
\def\Lyfive{\hbox{{\rm Ly}\kern 0.1em$5$}}
\def\Lysix{\hbox{{\rm Ly}\kern 0.1em$6$}}
\def\Lyseven{\hbox{{\rm Ly}\kern 0.1em$7$}}
\def\Lyeight{\hbox{{\rm Ly}\kern 0.1em$8$}}
\def\Lynine{\hbox{{\rm Ly}\kern 0.1em$9$}}
\def\Lyten{\hbox{{\rm Ly}\kern 0.1em$10$}}
\def\HeI{\hbox{{\rm He~}\kern 0.1em{\sc i}}}
\def\HeII{\hbox{{\rm He~}\kern 0.1em{\sc ii}}}
\def\FeI{\hbox{{\rm Fe~}\kern 0.1em{\sc i}}}
\def\FeII{\hbox{{\rm Fe~}\kern 0.1em{\sc ii}}}
\def\FeIII{\hbox{{\rm Fe~}\kern 0.1em{\sc iii}}}
\def\MnII{\hbox{{\rm Mn}\kern 0.1em{\sc ii}}}
\def\MgI{\hbox{{\rm Mg~}\kern 0.1em{\sc i}}}
\def\MgII{\hbox{{\rm Mg~}\kern 0.1em{\sc ii}}}
\def\MgIII{\hbox{{\rm Mg~}\kern 0.1em{\sc iii}}}
\def\MgIV{\hbox{{\rm Mg~}\kern 0.1em{\sc iv}}}
\def\MgX{\hbox{{\rm Mg~}\kern 0.1em{\sc x}}}
\def\NaI{\hbox{{\rm Na~}\kern 0.1em{\sc i}}}
\def\NV{\hbox{{\rm N~}\kern 0.1em{\sc v}}}
\def\NI{\hbox{{\rm N~}\kern 0.1em{\sc i}}}
\def\NII{\hbox{{\rm N~}\kern 0.1em{\sc ii}}}
\def\NIII{\hbox{{\rm N~}\kern 0.1em{\sc iii}}}
\def\OVI{\hbox{{\rm O~}\kern 0.1em{\sc vi}}}
\def\SiII{\hbox{{\rm Si~}\kern 0.1em{\sc ii}}}
\def\SiIII{\hbox{{\rm Si~}\kern 0.1em{\sc iii}}}
\def\SiIV{\hbox{{\rm Si~}\kern 0.1em{\sc iv}}}
\def\SII{\hbox{{\rm S~}\kern 0.1em{\sc ii}}}
\def\SIII{\hbox{{\rm S~}\kern 0.1em{\sc iii}}}
\def\SIV{\hbox{{\rm S~}\kern 0.1em{\sc iv}}}
\def\TiII{\hbox{{\rm Ti}\kern 0.1em{\sc ii}}}
\def\ZnII{\hbox{{\rm Zn}\kern 0.1em{\sc ii}}}
\def\kms{\hbox{km~s$^{-1}$}}
\def\cmsq{\hbox{cm$^{-2}$}}
\def\cc{\hbox{cm$^{-3}$}}

\def\apj{ApJ}%
\def\mnras{MNRAS}%
\def\aap{A\&A}%
\def\apjl{ApJ}
\def\aj{AJ}

\def\apjs{ApJS}

\def\pasp{PASP}
\def\nat{Natur}
\def\araa{AR\&A}

\newcommand {\apgt} {\ {\raise-.5ex\hbox{$\buildrel>\over\sim$}}\ }
\newcommand {\aplt} {\ {\raise-.5ex\hbox{$\buildrel<\over\sim$}}\ } 

\title[{}]{Detection of Two Intervening {\NeVIII} Absorbers Probing Warm Gas at $z\sim 0.6^{1}$}
\author[Pachat et al]
{
\parbox{\textwidth}{ 
Sachin Pachat$^{2}$ \thanks{E-mail:sachinpc@live.com},  
Anand Narayanan$^{2}$\thanks{Email: anand@iist.ac.in},
Vikram Khaire$^{3}$,
Blair D. Savage$^{6}$,
Sowgat Muzahid$^{5}$, 
and Bart P. Wakker$^{6}$
} 
\vspace*{10pt}\\ 
$^{1}$Based on observations with the NASA/ESA {\it Hubble Space Telescope}, obtained at the Space Telescope Science \\ Institute, which is operated by the Association of Universities for Research in Astronomy, Inc., under NASA contract NAS 05-26555\\
$^{2}$Indian Institute of Space Science \& Technology, Thiruvananthapuram 695 547, Kerala, INDIA\\  
$^{3}$National Centre for Radio Astrophysics, Tata Institute of Fundamental Research, Pune 411007, India \\ 
$^{5}$Leiden Observatory, Leiden University, P.O. Box 9513, 2300 RA Leiden, The Netherlands\\  
$^{6}$Department of Astronomy, The University of Wisconsin-Madison, 5534 Sterling Hall, 475 N. Charter Street, Madison WI 53706-1582, USA\\  
}   

\begin{document}
\date{}
\pagerange{\pageref{firstpage}--\pageref{lastpage}} \pubyear{2015}
\maketitle

\label{firstpage}

\vspace{20 mm}

\begin{abstract}

We report on the detection of two {\NeVIII} absorbers, at $z = 0.61907$ and $z = 0.57052$ in the $HST$/COS spectrum of background quasars  SDSS~J$080908.13+461925.6$ and SBS~$1122+594$ respectively. The {\NeVIII}~$770$ line is at $\sim 3\sigma$ significance. In both instances, the {\NeVIII} is found to be tracing gas with $T \gtrsim 10^5$~K, predominantly collisionally ionized, with moderate densities of $n_{\H} \lesssim 10^{-4}$~{\cc}, sub-solar metallicities and total hydrogen column densities of $N(\H) \gtrsim 10^{19}$~{\cmsq}. In the $z = 0.61907$ absorber, the low, intermediate ions and {\OVI} are consistent with origin in photoionized gas, with the {\OVI} potentially having some contribution from the warm collisional phase traced by {\NeVIII}. The $z = 0.57052$ system has {\HI} absorption in at least three kinematically distinct components, with one of them having $b(\HI) = 49~{\pm}~11$~{\kms}. The intermediate ionization lines, {\OVI} and {\NeVIII} are coincident in velocity with this component. Their different line widths suggest warm temperatures of $T = (0.5 - 1.5) \times 10^5$~K. Both absorbers are residing in regions where there are several luminous ($ \gtrsim L^*$) galaxies. The absorber at $z = 0.57052$ is within the virial radius of a $2.6L^*$ galaxy, possibly associated with shock heated circumgalactic material. 

\end{abstract}

\begin{keywords}
galaxies: halos, intergalactic medium, quasars: absorption lines, quasars: individual:SDSS J $080908.13+461925.6$, SBS~$1122+594$,  ultraviolet: general
\end{keywords}

\section{Introduction}\label{sec1}

Throughout cosmic history, $\gtrsim 90$\% of the baryons in the universe have resided outside of galaxies in the diffuse intergalactic and circumgalactic space. Compared to high redshifts ($z \gtrsim 1$) our understanding of the physical state of these baryons is less complete closer to the present epoch. At high redshifts, the photoionized intergalactic gas traced by the forest of {\Lya}  absorption lines seen in the spectrum of background quasars account for nearly all of the baryons in the universe. The {\Lya} forest corresponds to gas with densities of $n_{\H} \sim 10^{-5}$~{\cc} (over densities of, $\Delta = \rho/\bar{\rho} \sim 1 - 10$) and photoionization temperatures of $T \sim 10^4$~K \citep{Rauch1998,Cen1994,Escude1996,Rauch1997,Hernquist1996,Bolton2009}. These intergalactic baryons were kept ionized ($f_{\rm HI} = n_{\rm HI}/n_{\rm H} \sim 10^{-4}$), and their thermal state was maintained by the far-UV radiation background shaped by energetic photons escaping from AGNs and star-forming galaxies across the universe \citep{Escude1990,HM1996,Shull1999,HM2012}. 

With the formation of large scale structure, significant changes occurred in the phase structure of these intergalactic baryons. From a predominantly photoionized phase at $z \gtrsim 3$, the intergalactic gas evolved into a complex multiphase medium at low-$z$. Large scale numerical simulations of structure formation offer a clear insight into this process. The simulations, and also analytical estimates based on baryon measures in galaxies, clusters and the IGM, predict that as much as $50$\% of the baryons from the cool photoionized phase at high-$z$ transformed into a highly ionized plasma with temperatures of $T \sim 10^5 - 10^7$~K and densities of $n_{\H} \sim 10^{-4} - 10^{-5}$~{\cc} \citep{Persic1992,Cen1999,Dave2001,Valageas2002,Cen2006}. The phase change happened from shock-heating when baryonic matter, driven by gravitational forces, streamed into dark matter overdensity clumps where galaxies and clusters gradually formed. Simulations suggest these collisionally ionized \textit{warm-hot} baryons to be in regions with densities that are a factor of 1 - 1000 greater than the cosmic mean density of matter (\citealt{Cen1999}). At the low end, these overdensities represent the tenuous regions of the intergalactic medium and at the high end, they coincide with regions proximate to galaxies. Observing this diffuse \textit{warm-hot} plasma is not just important for completing the baryon census at low-$z$, but it can also offer valuable insights into the chemical abundances and multiphase properties of galaxy halos which are substantial reservoirs of baryons \citep{Maller2004,Fukugita2006}. 

Finding this shock-heated phase of gas has been a challenge because of its low density, high temperature and correspondingly low levels of gas neutral fraction ($f_{\HI} \sim 10^{-6} - 10^{-7}$). Absorption line studies of bright quasars in the ultraviolet have been the most successful method for probing this warm-hot diffuse gas. Among the possible absorption lines that are diagnostic of the low densities and high temperatures of the plasma, highly ionized metal species, mainly {\OVI}, and {\NeVIII}, have been in focus because of their relatively strong doublet transitions in the far-ultraviolet and their high ionization properties. 

The {\NeVIII} species is a tracer of collisionally ionized gas with $T \gtrsim 10^5$~K in the IGM as well as in galaxy halos. Hybrid models that simultaneously include collisional and photoionization processes associate {\NeVIII} with low density gas with $n_{\H} \sim 10^{-5}$~{\cc} at warm temperatures of T $\sim 5 \times 10^5$~K with collisions dominating the ionizations \citep{Narayanan2011a,Tepper-Garcia2013}. However, at those temperatures and moderate densities, the gas can radiatively cool rapidly within timescales of $t_{cool} < 10^8$~yrs to $T \sim 10^4$~K where photoionization dominates \citep{Hussain2017}. With the updated extragalactic background radiation given by \citet[KS15]{KS2015b}, \citet{Hussain2017} showed that the {\NeVIII} absorbers are most likely to be tracers of collisionally ionized gas if the gas metallicity is low ($Z < 0.1~Z_{\odot}$) such that the cooling is slow. Instead, if the gas metallicity is high ($Z \gtrsim Z_{\odot}$), there has to be some constant injection of mechanical energy into the system to counter the cooling due to the presence of metals. In other words, if the gas metallicity is near-solar or higher the {\NeVIII} absorber can be tracing cooler photoionized gas. Production of {\NeVIII} through photoionization alone can also happen in the presence of a hard radiation field as in the case of absorbers close to the central engines of quasars \citep{Petitjean1999,Ganguly2006,Muzahid2012}. 

The choice between photoionization and collisional ionization origin is much more divided for {\OVI} class of absorbers. They too can have a dual origin in cooler photoionized gas as well as warmer collisionally ionized medium \citep{Tripp2008,Oppenheimer2009,Tepper-Garcia2011}. In the latter case, the baryonic column density in the absorber is usually at least a factor of $10$ more than when the {\OVI} absorber is photoionized.  Quasar absorption line surveys have discovered more than 100 {\OVI} absorbers at low-$z$ \citep{Tripp2000, Tripp2008, Savage2002, Danforth2005, Thom2008, Thom2008b, Savage2014, Tumlinson2011, Danforth2016}. In a significant number of these cases, the {\OVIdblt} lines are found to be consistent with gas at temperature ranging from $10^5$ to $10^6$~K \citep{Tripp2000, Savage2002, Howk2002, Danforth2008, Lehner2009, Narayanan2010a, Narayanan2010b, Savage2011, Narayanan2012, Savage2014, Pachat2016}. Among these are also instances where the warm gas is directly evident (independent of ionization modeling) through the presence of thermally broad Lya absorption (BLA, b(HI) > 40 km/s) associated with the OVI.Table 5 in \citet{Savage2014} contains a list of 14 O VI systems where the associated BLAs imply  log T ranges from 5.0  to 6.14.   Their table also includes four additional O VI systems that have BLAs implying log T from 4.7 to 4.8.  Additional references to the literature are found in their detailed discussions of the properties of each system.. 

Compared to the fairly large frequency of incidence of {\OVI} systems, only ten intervening {\NeVIIIdblt} detections have been reported thus far \citep{savage2005, Savage2011, Narayanan2009, Narayanan2011a, Tripp2011, Narayanan2012, Meiring2013, Hussain2015, Bordoloi2016, Qu2016}. The low cosmic abundance of neon compared to oxygen ([Ne/O]$_\odot = -0.76$, \citealt{Asplund2009}) could be one of the reasons. The {\NeVIII} absorption appears at very low contrast against the background continuum for typical IGM column densities of $\log~[N(\HI), {\cmsq}] \sim 14 - 16$. The discovery of {\NeVIIIdblt} lines with adequate significance therefore requires high $S/N$ spectroscopic observations in the ultraviolet. 

Eight of the ten {\NeVIII} detections have come from $HST$/COS observations. The remaining two are based on data from the $FUSE$ satellite (see Table \ref{NeVIII_compile} for a summary). In all those instances, except \citet{Bordoloi2016} where ionization models are not discussed, the presence of {\NeVIII} was decisive in revealing the presence of warm collisionally ionized gas with $T \gtrsim 10^5$~K in an otherwise complex multiphase absorber \citep[however, see,][]{Hussain2017}. Moreover, in each case, the {\NeVIII} warm gas was found to be tracing a significant baryonic column of $N(\H) \gtrsim 10^{19}$~{\cmsq}. In several cases weak and broad {\Lya} absorption (BLA) associated with the {\NeVIII} gas was consistent with the high temperature estimate.
\begin{table*} 
\caption{Intervening {\NeVIII} detections so far}
\begin{center}  
\scriptsize
\begin{tabular}{lllllllllll}
\hline
 Line of sight               & $z_{abs}$ & Transition &$ W_r (m{\AA})$  &  $\log[N]$   & $\log[N(H)]$ & $log T$ K & $[X/H]$ & Origin of\\   
                             &           &            &                 &                     &                     &           & &{\NeVIII}\\
\hline
\\
HE 0226-4110                 & $0.2070$  &{\NeVIII} 770   & $ 32.9 \pm 10.5$ & $13.85^{+0.12}_{-0.17}$& $20.06$     &  $5.68 \pm 0.02$ &--- & Collisional  \\
\citep{savage2005}           &            &{\NeVIII} 780   & $24.9 \pm 10.6$  & $14.03^{+0.15}_{-0.24}$ & --- &--- & ---                 &  \\
\citep{Savage2011}           &            &{\NeVIII} 780   & $24.9 \pm 10.6$  & $14.03^{+0.15}_{-0.24}$ & --- &--- & ---            &  \\
                             &            &{\OVI}          &    ---           & $14.38 \pm 0.01$        & --- &--- & $-0.89 \pm 0.1 $  &  \\
\hline
 3C 263                      & $0.32566$ &{\NeVIII} 770   &$47 \pm 11.9$  & $ 13.98^{+0.10}_{-0.13}$  & ---     &  $5.80$ &--- & Collisional \\
\citep{Narayanan2009}        &           &                &               &                           &         &         &    &             \\

\hline
 PKS 0405-123                & $0.4951$  &{\NeVIII} 770   & $45 \pm 6 $   & $ 13.96 \pm 0.06$& $ 19.67 $      &  $5.70$ & $-0.6 \pm 0.3 $& Collisional\\
 \citep{Narayanan2011a} &           &{\NeVIII} 780   & $29 \pm 5 $   & $ 14.08 \pm 0.07$&   ---     &--- &$-0.6 \pm 0.3 $           &  \\
                             &           &{\OVI} 1038     & $153 \pm 5$   & $14.48 \pm 0.01 $&  ---      & ---&   ---        & \\
\hline
 PG1148+549                  & $0.6838$  &{\NeVIII} 770   &$51 \pm 12$    & $13.98 \pm 0.09$ & $19.80$   & $5.69$ & $>$ -0.5  &  Collisional\\
 \citep{Meiring2013}         &           &{\OVI} 1032     & $234 \pm 19$  & $14.47\pm 0.03$  &    ---       &  ---    &$>$ -0.5 &    \\   
                             &           &{\OVI} 1038     & $149 \pm 23$  & $14.50\pm 0.06$  &    ---       &  ---     &$>$ -0.5 &  \\
\\
                             & $0.7015$  &{\NeVIII} 770   & $28 \pm 5 $   & $13.75 \pm 0.07$&$19.00$    & $5.69$ &  $>$ -0.2 & Collisional \\
                             &           &{\NeVIII} 780   & $18 \pm 6$    & $ 13.86 \pm 0.11$    &   ---     &  ---    &$>$-0.2  &  \\
                             &           &{\OVI} 1032     & $168 \pm 19$  &  $14.29 \pm 0.04$    &   ---      & ---     &$>$-0.2  &  \\
                             &           &{\OVI} 1038     & $113 \pm 19$  &  $14.35 \pm 0.06$     &  ---      & ---     &$>$-0.2  &  \\
\\
                             & $0.7248$  &{\NeVIII} 770   & $26 \pm 8$    & $13.70 \pm 0.12$    &$ 18.90 $ & $5.72$ & $>$ 0    & Collisional \\
                             &           &{\NeVIII} 780   & $19 \pm 5$    & $13.87 \pm 0.10$     &   ---      &  ---    &$>$0  & \\
                             &           &{\OVI} 1032     & $71 \pm 20 $  & $13.84\pm 0.10$       &  ---     &   ---    &$>$0  &\\
                             &           &{\OVI} 1038     & $37\pm 12 $   & $13.87\pm 0.15$       &  ---     &  ---     &$>$0  &\\
\hline
 PG 1407+265                 & $0.5996$  &{\NeVIII} 780   & $52.8\pm 6.6$ & $14.15 \pm 0.18$  &$ 19.40 $ & $4.80$ & $>$ 0 & Photoionization \\
\citep{Hussain2015}          &           &{\OVI}          & ---           & $14.57 \pm 0.05$   &  ---     & ---    &--- &  \\    
                             
\hline
 LBQS 1435-0134              & $1.1912$  &{\NeVIII}       &      ---         & $13.96 \pm 0.17$      & $ 19.92 $ & --- &     0.3 & Collisional \\
\citep{Qu2016}                     &           &{\OVI}          & ---           & $14.49 \pm 0.05$      &     ---      &  ---   &  ---       &  \\
\hline
 PG 1206+459                 & $0.927$   &{\NeVIII} comp-1   & ---           & $13.71 \pm 0.29$     & $ 19.4 $ & $5.4$ &  3  & Collisional \\
\citep{Tripp2011}            &           &{\NeVIII} comp-2   & ---           & $14.08 \pm 0.08$     & $ 20.3 $ & $5.6$ & 1   &   \\
                             &           &{\NeVIII} comp-3   & ---           & $14.07 \pm 0.04$     &  ---     &  ---  & 0.5 &   \\
                             &           &{\NeVIII} comp-4   & ---           & $14.53 \pm 0.04$     &  $20.4 $ & $5.6$ &  1  &   \\
                             &           &{\NeVIII} comp-5   & ---           & $14.21 \pm 0.05$     &   ---    &  ---  & ---    &   \\
                             &           &{\NeVIII} comp-6   & ---           & $13.30 \pm 0.27$     &   ---    &  ---  & ---    &   \\
                             &           &{\NeVIII} comp-7   & ---           & $13.78 \pm 0.09$     &   ---    &  ---  & ---    &   \\
\hline
QSO J1154+4635               &           &{\NeVIII} 770      &    ---        & $14.65 \pm 0.08$     &   ---    &  ---  & ---  &   \\
\citep{Bordoloi2016}         &           &{\OVI}             &    ---        & $14.71 \pm 0.01$     &   ---    &  ---  & ---  &  \\
                            
\hline
SDSS~J$080908+$              & $0.61907$ &{\NeVIII} 770   & $28 \pm 8$    &  $ 13.76 \pm 0.14$   &$ 20.45$ & $>5.6$ & $<$ 0.5 & Collisional \\
$461925.6$ (this work)       &           &{\OVI} 1038     & $248 \pm 18$  &  $ 14.79 \pm 0.06$   &  ---       & ---    & --- &  \\
\hline
SBS~$1122+594$ (this work)              & $0.57052$ &{\NeVIII} 770   & $26 \pm  9$   &  $13.72 \pm 0.15$    &$ 19.70$ &$5.0$ &$<$ -0.8 & Collisional \\
                             &           &{\OVI} 1032     & $82 \pm 21$   &  $13.92 \pm 0.13$    &  ---       & ---    & -1.4 &  \\
\hline
\hline
\end{tabular}
\label{NeVIII_compile}
\end{center}
\begin{flushleft}
{Comments: The table is a compilation of all intervening {\NeVIII} detections so far. Column 4 is the equivalent width in the rest-frame of the absorber, column 5 is the apparent optical depth measured column densities, column 6 lists the total hydrogen column density (ionized and neutral combined), column 7 is the temperature reported by the authors for the {\NeVIII} gas phase, and column 8 is the dominant ionization mechanism for the production of {\NeVIII} as concluded by the respective authors (see \citealt{Hussain2017} for a remodeling of 7 these absorbers using the updated KS15 ionizing background, and the likelihood of a photoionization origin for the {\NeVIII} at supersolar metallicities).} 
\end{flushleft}
\end{table*} 
In this paper, we report on the discovery of two new {\NeVIII} absorbers at $z \sim 0.6$ towards the background quasars SDSS~J$080908.13+461925.6$ and SBS~$1122+594$. Though a detection is clearly evident in the $HST$/COS spectra, the formal significance of the {\NeVIII}~$770$ absorption feature is lower compared to some of the previous instances of {\NeVIII} detections. Detailed analysis of the absorption systems rule out a photoionization origin for the {\NeVIII} in either absorber. Using the revised extragalactic ionizing background of KS15, we find the {\OVI} to be consistent with a photoionized origin. 

In section \ref{sec2} we present information on the COS archival observations of both sightlines and the data analysis techniques used. In section \ref{sec3} and section \ref{sec4} we describe our measurements on the properties of both absorption systems, with emphasis on the {\NeVIII} lines. The ionization mechanisms, physical properties and chemical abundances of the absorber towards SDSS~J$080908.13+461925.6$ are discussed first in section \ref{sec5}, followed by the absorption system towards SBS~$1122+594$ in section \ref{sec6}. In section \ref{sec7}, we present information on the galaxies identified by the SDSS that are coincident in redshift with the absorbers. The main results are summarized in section \ref{sec9}, which is followed by a brief discussion on {\NeVIII} absorbers.

\begin{figure*}
\centering
\includegraphics[totalheight=0.85\textheight, trim=0cm 0cm -1cm 0cm, clip=true]{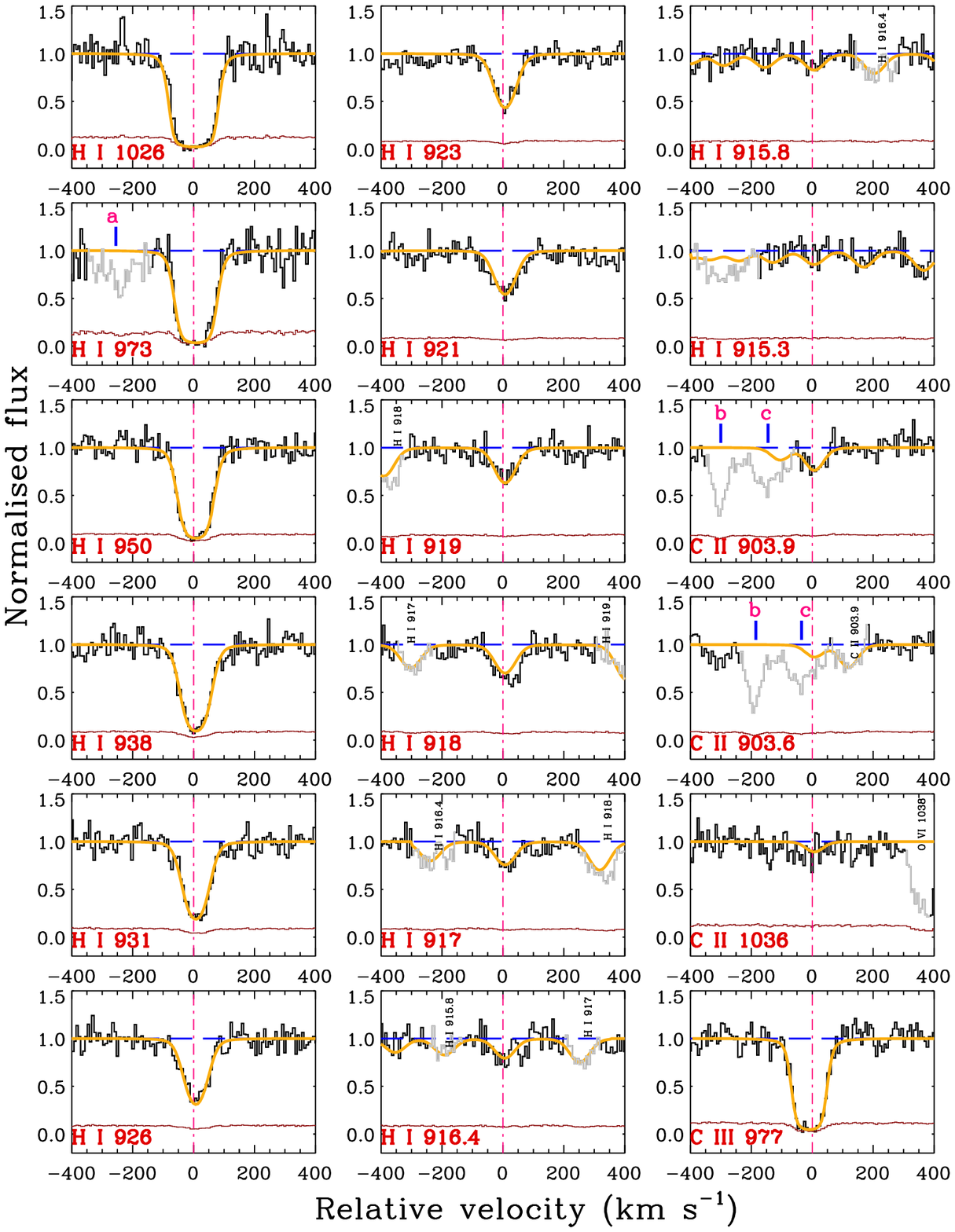}
\caption{System plot showing the important transitions against the rest-frame of the $z= 0.61907$ absorber towards the background quasar SDSS~$J080908.13+461925.6$. The absorption system covers a host of Lyman series lines, metal lines such as {\CII}, {\CIII}, {\NIII}, {\NIV}, {\OIII}, {\OIV}, {\OVI}, {\SIV} and {\SV}. The lines {\NII}~916, {\NII}~1084 and {\NeVIII}~780 are $3\sigma$ non detections. {\CIII}~$977$ is significantly saturated. Voigt profile fits are superimposed on the respective features as {\textit{yellow}} curves. The {\textit{gray}} regions indicate contaminations, i.e., absorption unrelated to this particular absorber. Contaminations are identified as (a) {\HI} 950 associated with the quasar, (b) possible {\Lya} from $z=0.2027$ and (c) {\OVI} 1032 at $z=0.4176$ confirmed by the detection of the corresponding {\OVI} 1038.}
\label{fig1}
\end{figure*}
\begin{figure*}
\centering
\includegraphics[totalheight=0.85\textheight, trim=0cm 0cm -1cm 0cm, clip=true]{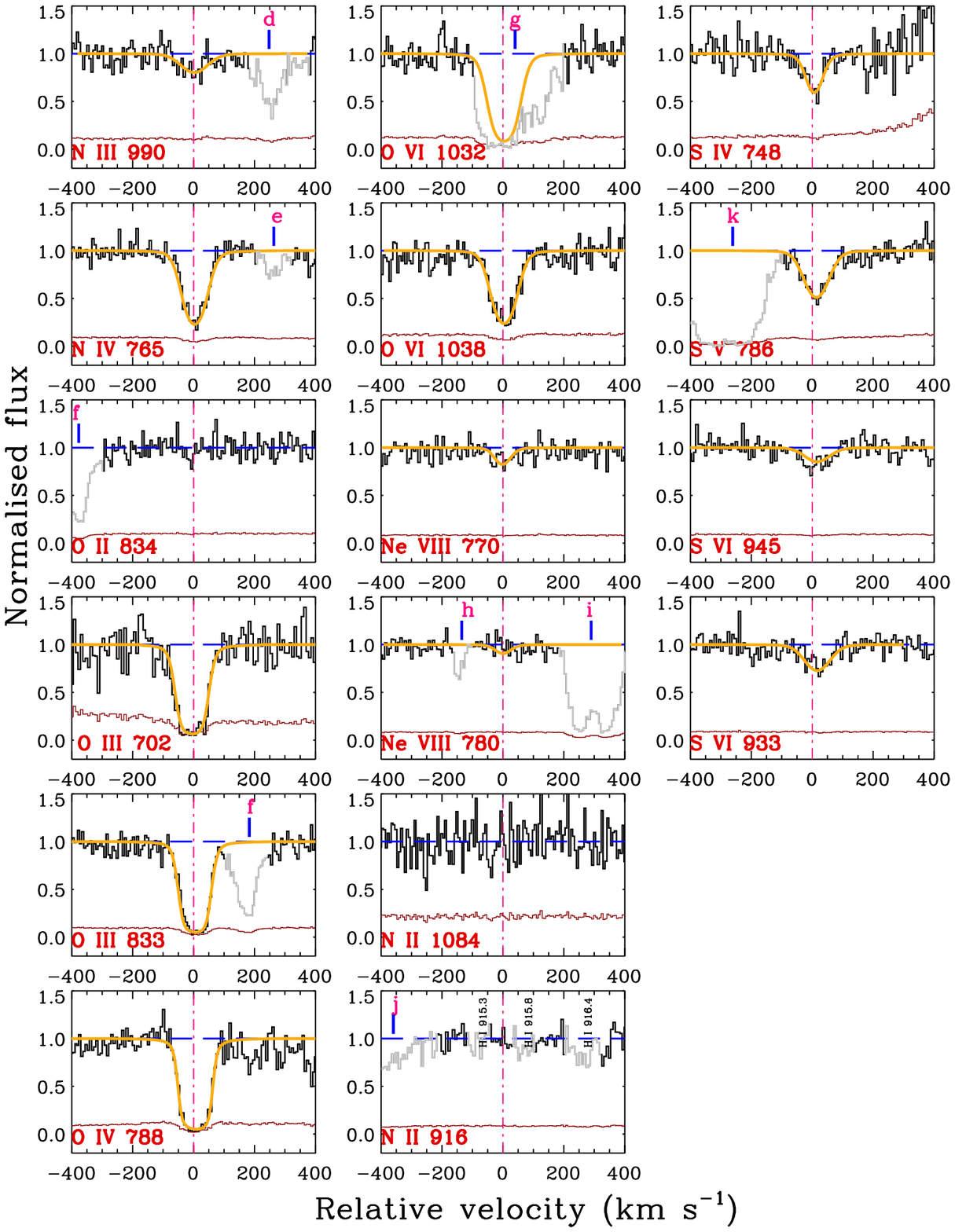}
\caption{\small{~Continuation of the system plot for the $z = 0.61907$ absorber shows single component Voigt profile fits to fairly well detected lines in the $z = 0.61907$ absorber.  The centroid of the  absorbing component is marked with a red vertical tick mark. The {\textit{gray}} regions in some of the panels indicate contamination. Voigt profile fit results are shown as \textit{yellow} curves, and the fit results are given in Table \ref{tab1}. The {\OIII} and {\OIV} lines are significantly saturated. The {\OVI}~1032 line is heavily blended with Galactic {\AlII}~$1671$. The expected {\OVI}~$1032$ absorption profile is shown in \textit{yellow} in that panel, based on the uncontaminated {\OVI}~$1038$ absorption. The contaminations in the various panels are (d) {\HI} 1026 at $z=0.5637$, for which the corresponding {\HI} 972 is detected, (e) possibly {\HI} 1216 at $z=0.020$ (f) {\OVI} 1032 at $z=0.3076$ confirmed by the presence of the weaker {\OVI} 1038 line, (g) Galactic {\AlII} 1671, (h) {\SiII} 1206 at $z=0.0466$ for which other metal lines are also detected (i) {\HI} 1216 at $z=0.0403$, (j) - and (k) possibly {\HI} 1216 at $z=0.0465$ respectively.}}
\end{figure*}

\section{The HST/COS Data}\label{sec2}
The SDSS~J$080908.13+461925.6~(z_{em} = 0.6587)$ quasar data presented here are far-UV medium resolution (instrumental FWHM $\sim 17$~{\kms}) spectra retrieved from the $Hubble~Space~Telescope$ / Cosmic Origins Spectrograph (COS) MAST archive\footnote{https://archive.stsci.edu/}. The instrument capabilities and in-flight performance are described in detail by \citet{Froning2009}, \citet{Dixon2010}, \citet{Osterman2011}, and \citet{Green2012}. The observations were carried out in 2010 as part of a COS dwarf galaxy halos project (PI. Jason Tumlinson, Prop ID: 12248, \citealt{Bordoloi2014}). The separate science exposures were processed using the STScI CalCOS (v2.17.3) pipeline. The data consists of G130M and G160M grating spectra of $4.9$~ks and $3.0$~ks integration times respectively. For each grating setting, there were multiple FP-position exposures. These individual science exposures were combined in flux units weighted by their exposure times using the coaddition routine developed by Charles Danforth\footnote{http://casa.colorado.edu/$\sim$danforth/science/cos/costools.html} and as described in \citet{Danforth2010}. The pipeline reduced COS spectra are over-sampled at 6 pixels per $17$~{\kms} resolution element. The spectra were therefore binned to the optimal Nyquist sampling of 2 pixels per resolution element. The resultant fully combined spectrum has a $S/N \sim 10 - 20$ per pixel over much of the wavelength range between $1135$~{\AA} and $1790$~{\AA}. The spectral resolution of COS is wavelength dependent with a maximum of $R = \lambda/\Delta \lambda \sim 20,000$ at near-UV wavelengths and decreasing monotonically to values of $R \sim 17,000$ at the edge of the far-UV covered by the G130M grating. The spectra were continuum normalized by fitting lower order polynomials across wavelength intervals of approximately $20$~{\AA}, avoiding regions containing absorption or strong emission features. 

The SBS~$1122+594$ $(z_{em} = 0.8514)$ COS individual exposures were also retrieved and processed in a similar manner. The quasar was observed for $9.8$~ks with the COS G130M grating and  $10.5$~ks with G160M grating settings. Under each grating setting, multiple exposures with different central wavelengths were used for covering the wavelength gap between the two detector segments of the instrument. The coadded spectrum has continuous wavelength coverage in the range $1135 - 1790$~{\AA}. The Nyquist sampled spectrum has $S/N$ varying in the range $5 - 15$~per pixel, with the peak $S/N$ occurring at $\lambda \sim 1200$~{\AA}, which is incidentally close to the redshifted location of {\NeVIIIdblt} lines in the absorber identified along this sight line. For both sightlines, our coadded version of the spectra agree with the independent extraction method of Bart Wakker, discussed in the appendix of \citet{Wakker2015}.

\section{The \lowercase{z} $=$ 0.61907 Absorber towards SDSS~J$080908.13+461925.6$}\label{sec3}
\begin{figure}
\centering
\includegraphics[totalheight=0.375\textheight, trim=0cm 0cm 0cm -1cm, clip=true]{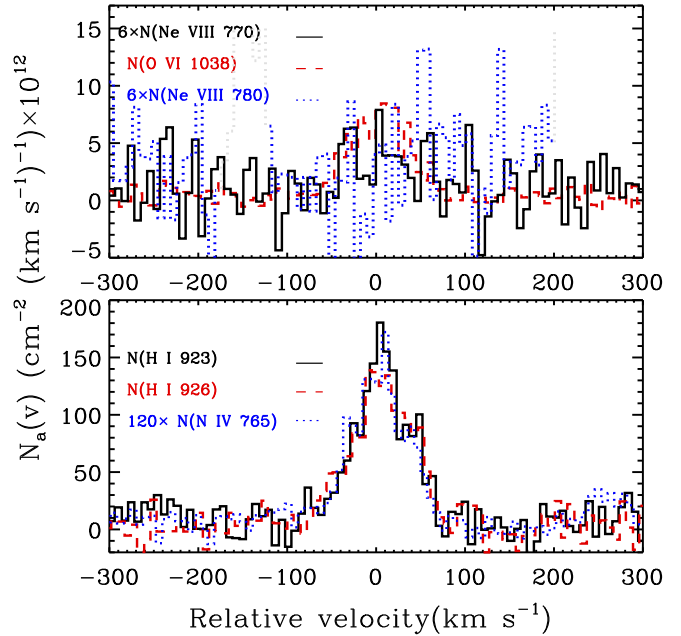}
\caption{Top panel is a comparison of the apparent column density profiles of {\NeVIII}~$770,~780$ and {\OVI}~$1038$ lines in the $z= 0.61907$ absorber. The apparent column density profile of {\NeVIII} 780 is truncated between [-180 -120]~{\kms} and $v > 200$~{\kms} where they suffer from contamination. The coincidence in velocity between {\NeVIII}~$770$ and {\OVI} lends support to the detection of {\NeVIII}~$770$.  In the bottom panel the apparent column densities of two of the unsaturated Lyman series lines and {\NIV}~$765$ are shown.}
\label{fig2}
\end{figure}

\ The continuum normalised regions of the COS spectrum covering important transitions of the absorber at $z = 0.61907$ are shown in Figure \ref{fig1}. The absorption system has {\HI} Lyman series ({\Lya}~--~{\HI}~$915$), {\CII}~$903.9$, $903.6$, {\CIII}~$977$, {\NIII}~$990$, {\NIV}~$765$, {\OIII}~$702, 833$, {\OIV}~$788$, {\OVI}~$1038$, {\SIV}~$748$, {\SV}~$786$, {\SVI}~$933, 945$ and {\NeVIII}~$770$ lines detected at $ \geq 3\sigma$ significance. In addition, the spectrum also covers {\OII}~$834$, {\NII}~$916$, and {\NeVIII}~$780$ which are non-detections. For the lines which are not detected, we quote the $3\sigma$ upper limit by integrating the spectrum through the same velocity range over which {\OVI} is detected.

\ The absorber is displaced from the systemic redshift of the quasar ($z_{em} = 0.6587$, \citealt{Hewett2010}) by $\Delta v = -7252$~{\kms} which is beyond the typical velocity offset cutoff of $\Delta v \lesssim 5000$~{\kms} used to differentiate associated absorbers from intervening systems e.gs., \citep{Foltz1986,Shen2012}. To rule out the possibility of the absorber being associated with the AGN, we compared the apparent column density profiles of higher order Lyman series lines (see Figure \ref{fig2}). The lines show similar apparent column density across the full range of the velocity of absorption, implying little partial coverage of the background AGN continuum. More generally, the high ionization lines associated with outflows are expected to be significantly stronger and saturated than the absorption in {\NeVIII}, {\OVI}, or the unsaturated absorption in {\SV} or {\SVI} seen for this system \citep{Muzahid2013,Fox2008}. Based on the absence of any observational signatures, we conclude that the absorption is not tracing quasar-driven outflows or gas close to the AGN central engine. 

\ The Voigt-profile fitting software VPFIT (ver 10.0)\footnote{http://www.ast.cam.ac.uk/$\sim$rfc/vpfit.html} was used to estimate the column density, Doppler $b$-parameter and the rest-frame velocity centroids of the lines.  The atomic line list and oscillator strength values used for fitting are from \citet{morton2003} for $\lambda > 912$~{\AA} and from \citet{verner1996} for ${\lambda} \leq 912$~{\AA}. The atomic data for {\NeVIIIdblt} are $\lambda = 770.4089,~780.3240$~{\AA}, and $f_{770} = 0.1030,~f_{780} = 0.0505$ respectively \citep{verner1996}. The spectral resolution of COS is known to be wavelength dependent. Therefore, while fitting, we used the empirical line spread functions developed by \cite{Kriss2011} for the respective COS gratings. The synthetic Voigt profiles were convolved with the COS line spread function nearest in wavelength to the redshifted location of the absorption line. Profile fit results are shown in Table \ref{tab1}. The line profile models are shown in Figure \ref{fig1}. The errors listed from the profile fit are a combination of statistical errors and continuum placement errors.  The latter quantity was estimated by exploring a few different continuum fits and the resultant range of values obtained for column densities and $b$-values by fitting the region. Overall, we found the statistical noise in the data to be more significant than the systematic errors. 
\begin{table*} 
\caption{Line measurements for the absorber at $z=0.61907$ towards SDSS~J$080908.13+461925.6$} 
\begin{center} 
\renewcommand{\arraystretch}{0.9} 
\begin{tabular}{lrccc}
\hline 
\multicolumn{5}{c}{Voigt Profile Measurements} \\ 
\hline 
\hline
    Transition          & $v$ (\kms)     &  $b$(\kms)       &  log~$[N~(\cmsq)]$ & Total log~$[N~(\cmsq)]$   \\   
\hline
{\NIV} 765                      & $ -30 \pm 4 $    & $14 \pm 4 $  & $ 13.47 \pm 0.11$  &                                \\  
 (3 comp)                       & $   6 \pm 2 $    & $15 \pm 5 $  & $ 13.85 \pm 0.06$  &                                \\          
                                & $  43 \pm 3 $    & $11 \pm 3 $  & $ 13.37 \pm 0.07$  & $14.09 \pm 0.06$               \\        
\\
{\NIV} 765(single comp)         & $ 3 \pm 2 $    & $41 \pm 2 $  & $ 14.05 \pm 0.02$  &                                \\ 
\\  
{\HI} 919-1026                  &  $-31  $         &  $30 \pm 9 $ & $ 15.50 \pm 0.08 $ &                                \\    
   (3 comp)                     &  $  5  $         &  $18 \pm 9 $ & $ 15.82 \pm 0.05 $ &                                \\              
                                &  $ 44  $         &  $15 \pm 4 $ & $ 15.52 \pm 0.06 $ & $ 16.10 \pm 0.05$              \\         
\\
{\HI} 919-1026  (single comp)   &  $-4  $          &  $41 $       & $ 16.13 \pm 0.02 $ &                                \\    
\\
{\OVI} 1038                     & $ 4  \pm 2 $     & $44 \pm 3 $  & $14.88\pm 0.20 $   & $  14.88 \pm 0.20 $             \\
{\OIV} 788                      & $ 6 \pm 1  $     & $24 \pm 4 $  & $16.53 \pm 0.70^{1}$   & $16.53 \pm 0.70^{1}$                  \\ 
{\OIII}                         & $ 4 \pm 2   $    & $ 32 \pm 3 $ & $ 15.52 \pm 0.15^{1}$  & $ 15.52 \pm 0.15^{1}$                \\
{\CIII} 977                     & $-10 \pm 1 $     & $ 33 \pm 3$  &  $14.71 \pm 0.20^{1}$  & $  14.71 \pm 0.20^{1}$               \\
{\CII} 903.9                    &  $ 8 \pm 6 $     & $ 37 \pm 9$  & $13.42 \pm 0.10 $  & $  13.42 \pm 0.10 $             \\
{\SIV} 748                      & $ 7 \pm 3 $      &  $ 32 \pm 5 $ & $ 13.59 \pm 0.05 $ & $ 13.59 \pm 0.05 $              \\
{\SV} 786                       & $ 13 \pm 3 $     &  $ 47 \pm 4 $ & $ 13.35 \pm 0.03 $ & $ 13.35 \pm 0.03 $              \\
{\NeVIII} 770                   & $ 11 \pm 12$     &  $69 \pm 20 $ & $ 13.96 \pm 0.10$  & $ 13.96 \pm 0.10$               \\
{\CII}  903.6 -1036             & $7 \pm  6$       &  $36 \pm 9$   & $ 13.42 \pm 0.07$  & $ 13.42 \pm 0.07$  \\
{\SVI} 934                      & $17 \pm 5$       &  $52 \pm 7$   & $ 13.49 \pm 0.05$  &  $ 13.49 \pm 0.05$ \\
{\NIII} 990                     & $-4 \pm 9 $      &  $51 \pm 16$   & $13.83 \pm 0.10$  &  $13.83 \pm 0.10$ \\
\hline
\multicolumn{5}{c}{Integrated Apparent Optical Depth Measurements} \\ 
\hline
\hline
    Transition     &    $W_r (m{\AA})$ &                          & log~$[N~(\cmsq)]$   & [-v,v]({\kms}) \\   
\hline
{\OVI} 1038    & $248 \pm 18$    &                                & $ 14.79 \pm 0.06$    &   [-85, +77]\\
\\
{\NIV} 765     & $182 \pm 10$    &                                & $ 13.95 \pm 0.05$    &   [-85, +77] \\
\\
{\OIV} 788     & $276 \pm 11 $   &                                & $ > 15.0 $           &   [-85, +77] \\
\\
{\NIII} 990    & $58 \pm 17 $    &                                & $13.78 \pm 0.15$     &   [-85, +77] \\ 
\\
{\CIII} 977    & $393 \pm 12$    &                                 &  $  >14.1$           &    [-75,50]   \\
\\
{\OIII} 833    & $281 \pm 10 $   &                                 &   $ > 15.0 $         &    [-85, +77] \\ 
{\OIII} 702    & $267 \pm 16 $   &                                 &   $ > 15.0 $         &    [-85, +77] \\
\\
{\CII} 1036    & $<63$           &                                 & $  <13.8$             &   [-85, 77]  \\ 
{\CII} 904.9    & $67 \pm 12$     &                                 & $ 13.50 \pm 0.08$     &   [-85, 77]  \\
{\CII} 904.6    & $<140$          &                                 & $ <14.2$              &   [-85, 77]  \\
\\
{\SIV} 748     & $80 \pm 14 $    &                                 &  $ 13.59 \pm 0.09$    &   [-85, +77]  \\
{\SIV} 809     & $46 \pm 13 $    &                                 &  $ 13.93 \pm 0.12$    &   [-85, +77]  \\
\\
{\SV} 786      & $110 \pm 10 $   &                                 & $ 13.25 \pm 0.06 $    &   [-85, +77]  \\
\\
{\SVI} 945     & $< 66 $         &                                 & $  < 13.63$           &   [-85, +77]  \\
{\SVI} 934     & $60 \pm 10$     &                                 & $ 13.30 \pm 0.08$     &   [-85, +77]  \\
\\
{\NeVIII} 770  & $28 \pm 8$      &                                 & $ 13.76 \pm 0.14$     &   [-85, +77]  \\
\\          
{\HI} 1026     & $> 515$         &                                 & $> 15.2$              &    [-105,110] \\
{\HI} 972      & $> 423$         &                                 & $> 15.6$              &    [-105,110] \\
{\HI} 950      & $> 365$         &                                 & $> 15.9$              &    [-105,110] \\
{\HI} 938      & $317 \pm 12$    &                                 & $ 16.00 \pm 0.05 $    &    [-105,110] \\
{\HI} 931      & $266 \pm 13$    &                                 & $ 16.08 \pm 0.06 $    &    [-105,110] \\
{\HI} 926      & $213 \pm 13$    &                                 & $ 16.08 \pm 0.05 $    &    [-105,110] \\
{\HI} 923      & $126 \pm 13$    &                                 & $ 16.14 \pm 0.06 $    &    [-100,100] \\
{\HI} 921      & $134 \pm 14$    &                                 & $ 16.13 \pm 0.06 $    &    [-100,100] \\
{\HI} 919      & $ 73 \pm 11$    &                                 & $ 15.97 \pm 0.09 $    &    [-60,75]   \\
{\HI} 918      & $100 \pm 13$    &                                 & $ 16.25 \pm 0.09 $    &    [-60,75]   \\   
{\HI} 917      & $ 50 \pm 12$    &                                 & $ 16.03 \pm 0.11 $    &    [-60,75]   \\   
{\HI} 916.4    & $ 55 \pm 12$    &                                 & $ 16.15 \pm 0.11 $    &    [-60,75]   \\ 
\hline
\hline
\end{tabular}
\label{tab1}
\end{center}
\begin{flushleft}
{Comments:~. Measurments on the various lines associated with the $z=0.61907$ towards SDSS~J$080908.13+461925.6$. The measurments were done using the apparent optical depth (AOD) and Voigt profile fitting techniques. The doublet / multiplet lines were separately and simultaneously fitted. The various columns list the rest-frame equivalent width, centroid velocity of the absorbing components, their $b$ parameters, column densities, and the velocity ranges of integration used to estimate these parameters. Among the metal lines, {\CIII}, {\OIII} and {\OIV} are strongly saturated. The formal errors given by the profile fitting routine, listed in the table, are too small and do not account for the uncertainty due to the saturation of the line feature. In the ionization models, we use the AOD column density measurements on these lines as lower limits.}
\end{flushleft}
\end{table*}
\ We have relied on single component profiles to model the absorption seen in {\HI} and the various metal lines. A few of the Lyman series lines (particularly {\HI}~$923$ and {\HI}~$926$) suggest multi-component structure to the core absorption in {\HI}, albeit at low significance. The {\NIV}~$765$ line also suggests possible kinematic sub-structure. The apparent column density profiles of these lines are compared in Figure \ref{fig2}. A free-fit to the {\NIV} line recovers three components at $v \sim -30, +6, +43$~{\kms}. A similar three component structure is not visibly evident in the unsaturated higher order Lyman lines, except {\HI}~$923$ and {\HI}~$926$, although there appears to be complexity to the {\HI} velocity structure. 

\ Taking a cue from the component structure seen in {\NIV}, we attempted a simultaneous three component fit to the Lyman series lines to determine the {\HI} corresponding to the components of {\NIV}. The fit results are given in Table \ref{tab1}. The column density profiles of the various Lyman series lines slightly differ from each other due to statistical fluctuations in the noise spectrum. Because of this, the multi-component profile models do not fit all the Lyman series lines equally well. The errors on column densities and $b$-parameters derived from the simultaneous fit reflect this. In Sec 5.2.1, we discuss how the ambiguity in the component structure for {\HI} affects the elemental abundance estimate for a warm plasma traced by the {\NeVIII} in this absorber. 

\ No unique sub-component structure is evident in the unsaturated {\SIV}~$748$, {\SV}~786, {\SVI}~$933, 945$ lines, and hence these lines are fitted with a single component. The comparatively strong {\OIII}, {\CIII} and {\OIV} lines are saturated. We fit these lines also with single components. The true error in the profile fit results for such saturated lines are likely to be larger than what the fitting routine suggests.  

\ Using the apparent optical depth (AOD) method of \cite{savage1991}, we estimated the total column density of the various ions by integrating over the full velocity range over which each line feature is seen. Using this method we also derived upper limits on column densities for lines which are non-detections and lower limits for the lines which are heavily saturated. These values are also tabulated in Table \ref{tab1}. 

\ The COS spectrum provides information on {\OII}, {\OIII}, {\OIV} and {\OVI}. The {\OII} is a non-detection, whereas the {\OIII} and {\OIV} lines are strong and quite probably saturated. The similarities in the profile structure of {\OIII} and {\OIV} suggest a possible origin in the same gas phase. Single component Voigt profile models explain fairly well the absorption in {\OIV} and also the doublet lines of {\OIII}, with $\chi^{2}$ goodness of fit values close to one. However, the profile fit results for these intermediate ionization lines are not unique. We find line saturation effects dominating the uncertainty in the profile fits. As a result, models with different combinations of $N$ and $b$ can explain the observed absorption in {\OIII} and {\OIV} without compromising on the quality of the fit. We therefore adopt the integrated apparent column density lower limits of $\log~N_a(\OIII) > 15.0$ and $\log~N_a(\OIV) > 15.0$ for subsequent analysis. In the case of {\CIII} also we use the $\log~N_a(\CIII) > 14.1$, for similar reasons. The {\CII}~$903.6$ line suffers from blending with {\OVI} 1032 from an absorber at $z = 0.4176$ . We use the {\CII}~$903.9$ line to constrain the {\CII} column density. 

\ The {\OVI}~$1032$ line is heavily blended with {\AlII}~$1671$ absorption from the local interstellar medium. The {\OVI}~$1038$ is, however, a clean feature, which we use for constraining the column density and $b$-parameter. A single component fit to the {\OVI}~$1038$ yields $\log N(\OVI) = 14.88\pm 0.20$ and $b(\OVI) = 44 \pm 3$ which is comparable to the integrated apparent optical depth column density of $\log N(\OVI) = 14.79~{\pm}~0.06$ to within their $1\sigma$ uncertainty range. This implies that the {\OVI}~$1038$ is adequately resolved by COS. From the $N$ and $b$-values measured from {\OVI}~$1038$, we synthesized the {\OVI}~$1032$ line profile. The synthetic profile, when superimposed on the data reveals the extent of contamination from Galactic {\AlII} (see the model profile in the {\OVI}~$1032$ panel of Figure \ref{fig1}), which is significant and difficult to correct for. 

\begin{figure*}
\centering
\includegraphics[totalheight=0.85\textheight, trim=0cm 0cm 0cm 0cm, clip=true]{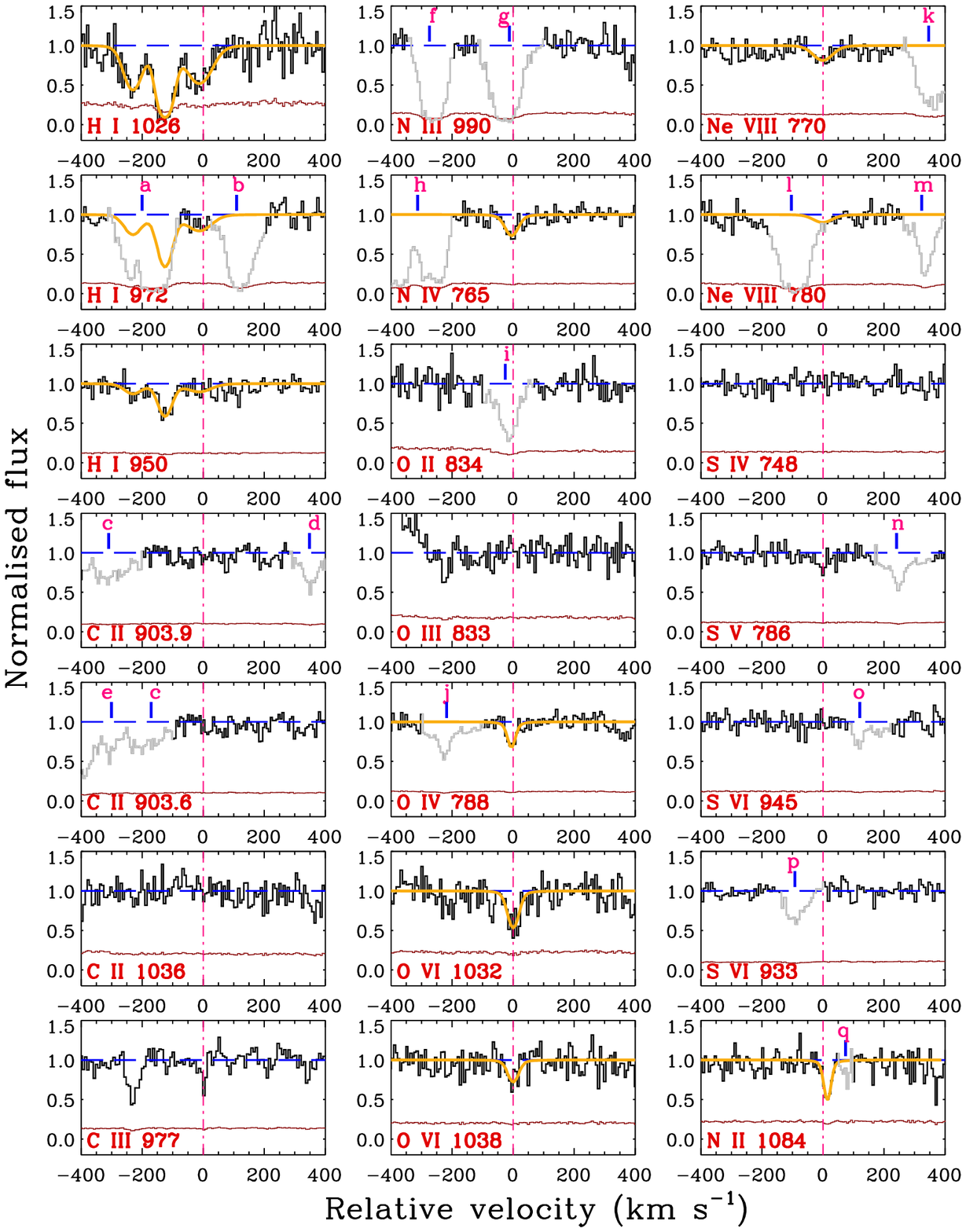}
\caption{The continuum normalised spectra of the $ z= 0.57052$ absorber towards SBS~$1122+594$. The {\textit{gray}} regions in the various panels indicate contamination, i.e., absorption not associated with this system. Some of the major contaminations are (a) and (b) Galactic {\SiII}~$1527$, (c), (d), and (e) associated {\NeVIII}~$770$, (f) {\Lya} at $z = 0.2777$ for which corresponding {\Lyb} and Ly$\gamma$ are identified, (g) {\CIV}~$1548$ at $z=0.0040$ for which corresponding {\Lya}, {\SiIVdblt}, {\SiIII} lines are present (h) Galactic {\NI}~$1200$, (i) {\Lyb} at $z = 0.2777$ for which corresponding {\Lya} and Ly$\gamma$ are present, (j) possibly {\Lya} at $z = 0.0169$, (k) {\SiIII}~$1206$ at $z = 0.0040$ confirmed by the presence of corresponding {\Lya}, {\CIVdblt}, {\SiIVdblt} lines (l) Ly$\beta$ at $z=0.1944$ for which other Lyman series lines, {\OVIdblt} etc are identified (m) possibly {\Lya} at $z = 0.0093$, (n) possibly {\Lya} at $z = 0.0163$, (o) possibly {\Lya} at $z=0.2207$, and (p) {\OVI}~$1032$ at $z = 0.4201$ for which higher order Lyman series lines are identified.}
\label{fig4}
\end{figure*}

\subsection{The {\NeVIIIdblt} Detection}

\ The {\NeVIIIdblt} is a very weak feature in the $z=0.61907$ absorber. The ion is detected with $\gtrsim 3\sigma$ significance only in the stronger member of the doublet. The transitions fall at the blue end of the G130M grating exposures. The three G130M integrations add to $3.1$~ks of exposure time at the redshifted wavelength location of {\NeVIIIdblt} lines. The exposures were obtained with two different G130M central wavelength settings (one exposure centered at $1291$~{\AA} and two exposures at $1309$~{\AA}) which result in the dispersed light getting shifted in the detector space. This helps to reduce the detector fixed pattern noise features during coaddition. 

The {\NeVIII}~$770$ line has a rest-frame equivalent width of $W_r = 28~{\pm}~8$~m{\AA} when integrated over [-85, 77]~{\kms}. This is the same velocity range over which absorption from {\CIII}, {\OIII}, {\OIV}, and {\OVI}~$1038$ are detected. The {\NeVIII}~$770$ feature is thus detected with a significance of $3.5 \sigma$. The uncertainty quoted for the equivalent width is inclusive of statistical and continuum placement errors. A more stringent estimate on the detection significance can be arrived at by including a systematic uncertainty of $\sim 10$~m{\AA} from residual fixed pattern noise features that could be present in COS data \citep{Savage2014}. This would bring the {\NeVIII}~$770$ significance down to $2.2\sigma$. The non-detection of {\NeVIII}~$780$ at $\geq 3 \sigma$ is consistent with the expected $2:1$ equivalent width ratio between the two lines of the doublet. 

\begin{figure*}
\centering
\includegraphics[width=1.0\textwidth, trim=0cm 0cm -1cm -1cm, clip=true]{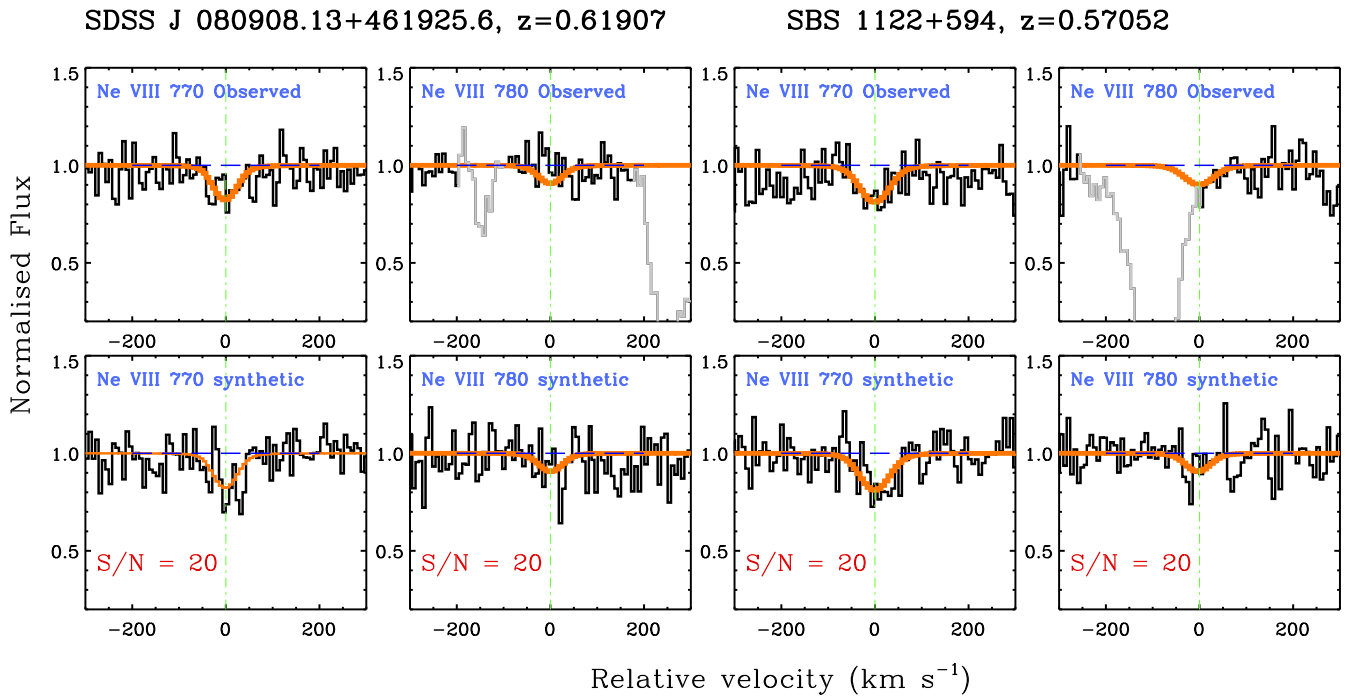}
\caption{The top panel of the first column shows the observed {\NeVIII}~$770$ in the $z = 0.61907$ absorber, with the Voigt profile model superimposed. The bottom panel is a synthetic {\NeVIII}~$770$ feature with identical $N$ and $b$ as the observed line at $S/N = 20$. The top and bottom panels of the second column show the corresponding observed and synthetic {\NeVIII}~$780$ spectra respectively. The synthetic spectra suggests that the weak {\NeVIII} can be a non-detection in the $780$~{\AA} line at the $S/N$ of the data, in agreement with observations. A similar analysis for the $z=0.57052$ absorber is shown in panels in the third and fourth column.}
\label{NeVIII_model}
\end{figure*}

The non-detection of {\NeVIII}~$780$ prompted us to investigate the validity of the {\NeVIII} detection in greater detail. We synthesized {\NeVIIIdblt} lines by convolving a model absorption feature (of $N$, and $b$-value obtained from fitting the {\NeVIII}~$770$) with a Gaussian kernel of FWHM = $17$~{\kms} (resolution of COS). Poisson noise was added to this synthetic profile to simulate $S/N = 20$ per wavelength bin ($1/2$ a resolution element). This approximately matches the $S/N$ of the data in the region where the {\NeVIII} occurs. The result of this exercise, shown in Figure \ref{NeVIII_model}, is consistent with the low significance of {\NeVIII}~$770$ and the non-detection of the {\NeVIII}~$780$ lines.

In the independent line identifications for this sightline done by \citet{Danforth2016} and one of the co-authors (Bart Wakker), the {\NeVIII}~$770$ absorption at $\lambda = 1247.35$~{\AA} is not identified with any line associated with other absorbers along this sightline. Nonetheless, we cannot fully eliminate the possibility of the {\NeVIII}~$770$ absorption being a weak low redshift {\Lya} interloper.   

We measured the strength of the {\NeVIII}~$770$ in the separate G130M integrations as well. The absorption feature is detected with an equivalent width of $W_r = 28~{\pm}~11$~m{\AA} in the exposure with grating central wavelength of $\lambda_c = 1291$~{\AA}, and $W_r = 51~{\pm}~20$~m{\AA} and $W_r = 36~{\pm}~20$~m{\AA} in the two exposures with central wavelengths of $\lambda_c = 1309$~{\AA}. The feature thus has a mean detection significance of $2.3\sigma$ in the individual science exposures. 

We made a closer investigation to find out whether fixed pattern noise or similar instrumental artefacts are affecting our measurement in anyway. Since two out of the three integrations of the G130M grating were carried out with the same central wavelength and FP-SPLIT position set-up of the grating, we cannot compare the individual exposures to know whether a fixed pattern feature is occurring at the {\NeVIII}~$770$~{\AA} redshifted wavelength of $1247.35$~{\AA}. Instead, we looked at this detector space in five other quasar observations done approximately in the same period with identical grating settings and found no evidence for any instrumental contamination. 

In Figure \ref{fig2}, we compare the apparent column density profiles of {\NeVIII} and {\OVI}. Though the {\NeVIII} line is much weaker than {\OVI}, in the approximate velocity range of $[-100, 100]$~{\kms} the absorption seen in {\OVI} is well matched by the absorption in {\NeVIII}. The similarity in the kinematics lends further support to the identification of the {\NeVIII}~$770$ line. 

The detection significance of {\NeVIII}~$770$ line is lower than the significance of most of the COS {\NeVIII} detections reported thus far \citep{Narayanan2009, Narayanan2011a,Tripp2011,Narayanan2012, Meiring2013,Hussain2015} but is similar to the first {\NeVIII} detection in the IGM reported by \citet{savage2005}, where the $770$ line was detected with a significance of $3.1 \sigma$. However, in the {\NeVIII} detection reported by \citet{savage2005}, the weaker $780$ line was also a formal detection with $2.3\sigma$ significance, resulting in a higher detection significance of $3.9 \sigma$ jointly for the lines of the doublet. 

\section{The $\lowercase{z}=0.57052$ Absorber towards SBS~$1122+594$}\label{sec4}

The absorber is detected in {\HI}, {\OIV}, {\NIV}, {\OVI} and {\NeVIII}. Continuum normalized velocity plots of important transitions are shown in Figure \ref{fig4}. The {\HI} clearly shows three distinct components at at $-8$, $-122$, $-227$~{\kms}. The metal lines are all aligned with the $-8$~{\kms} component, which interestingly is not the strongest component in {\HI}. The COS spectrum also covers wavelength regions where lines from {\CII}, {\CIII}, {\OIII}, {\SIV}, and {\SVI} are expected. The line measurements are given in Table \ref{tab2}. The {\OVIdblt} lines were simultaneously fitted freely with Voigt profiles. The lines fall in a comparatively low $S/N$ ($\sim 6$ per $17$~{\kms} resolution element) part of the spectrum, which is reflected in the large uncertainty associated with the fit parameters, and in the unusual equivalent width ratio between the two lines of the doublet. The doublet lines have a combined detection significance of $4.1 \sigma$. The column density from profile fitting is similar to the integrated apparent column density obtained for the $1032$~{\AA} and $1038$~{\AA} lines. This indicates little unresolved saturated structure in the {\OVI} profiles. The same is also true for {\Lyb}, {\NIV}~$765$ and {\OIV}~$788$ transitions.

\begin{table*} 
\caption{Line measurements for the absorber at $z=0.57052$ towards SBS~$1122+594$} 
\begin{center}  
\begin{tabular}{lccc}
\hline
\multicolumn{4}{c}{Voigt Profile Measurements} \\   
\hline
    Transition     &                $v$ (\kms)     &  $b$(\kms)       &  log~$[N~(\cmsq)]$        \\   
\hline  
{\HI} 950-1026 &                  $   -8   \pm 7 $ & $49 \pm 11 $ & $14.46 \pm 0.07$    \\
               &                  $ -122   \pm 2 $ & $26 \pm  3 $ & $14.97 \pm 0.04$    \\
               &                  $ -227   \pm 5 $ & $34 \pm  7 $ & $14.44 \pm 0.06$    \\  
\\
{\OVI} 1032-1038&                $   4 \pm 5  $   & $20 \pm 6 $  & $13.95 \pm 0.06$    \\
\\
{\OIV} 788     &                     $ 0 \pm 4 $   & $ 12 \pm 6 $ & $13.84 \pm 0.09$    \\
\\ 
{\NIV} 765     &                     $0 \pm 5 $    & $25 \pm 7  $ & $13.15 \pm 0.07$    \\
\\
{\NeVIII} 770  &                      $-2 \pm  10$ &  38 $\pm 8$    &$13.92 \pm 0.14$      \\ 
\\
{\NII} 1084    &                      $10 \pm 2$    & 10 $\pm$ 4     &$13.93 \pm 0.09$     \\
\hline
\multicolumn{4}{c}{Integrated Apparent Optical Depth Measurements} \\ 
\hline
    Transition     &  $W_r (m{\AA})$   &  log~$[N~(\cmsq)]$ & [-v,v](\kms) \\ 
\hline  
{\NeVIII} 770  &  $26 \pm  9$                  & $13.72 \pm 0.15$ &  [-60, 40] \\     
{\NeVIII} 780  &  $< 22     $                 &$    < 13.9  $    &  [-15, 40] \\      
{\CIII} 977    &  $< 40     $                  &$  < 13.0    $    &  [-60, 50] \\
{\OIII} 833    &  $< 45     $                 &$  < 13.8    $    &  [-60, 50] \\ 
{\SVI} 933     &  $< 33     $                 &$  < 13.0    $    &  [-60, 50] \\  
{\SVI} 945     &  $< 36     $                  &$  < 13.3    $    &  [-60, 50] \\ 
{\SIV} 748     &  $< 30     $                 &$< 13.1      $    &  [-60, 50] \\    
{\CII} 903.9   &  $< 30     $                  &$< 13.0      $    &  [-60, 50] \\  
{\CII} 903.6   &  $< 30     $                   &$< 13.3      $    &  [-60, 50] \\   
{\CII} 1036    &  $< 60     $                  &$< 13.7      $    &  [-60, 50] \\   
{\NII} 1084    &  $52 \pm 17$                  &$13.75 \pm 0.18$  &  [-40, 20] \\  
\\
{\HI} 1026     &  $ 142 \pm 23 $              & $14.40 \pm 0.14$ &  [-60, 50] \\
{\HI} 972      &  $  45 \pm 13 $              & $14.31 \pm 0.13$ &  [-60, 50] \\
{\HI} 950      &  $  32 \pm 13 $              & $14.47 \pm 0.14$ &  [-60, 50] \\   
\\
{\OVI} 1032    & $82 \pm 21$                  & $13.92 \pm 0.13$ &  [-60, 50] \\ 
{\OVI} 1038    & $25 \pm 19$                 & $  < 14.0      $ &  [-60, 50] \\ 
\\
{\OIV} 788     & $25 \pm 9 $                   & $13.66 \pm 0.15$ &  [-40, 10] \\  
\\
{\NIV} 765     & $41 \pm 10$                   & $13.16 \pm 0.12$ &  [-60, 50] \\ 
\hline
\hline
\end{tabular}
\label{tab2}
\end{center}
\begin{flushleft}
\footnotesize{Comments:~The line measurments were done using apparent optical depth and Voigt profile fitting techniques. The various columns list the rest-frame equivalent width, centroid velocity of the absorbing components, their $b$ parameters, column densities, and the velocity range of integration used to estimate these parameters.}
\end{flushleft}
\end{table*}
The {\Lyb} and {\Lyd} lines were also simultaneously fitted with three components. The {\Lyg} transition was excluded from the profile fitting as the region is severely contaminated by Galactic {\SiII}~$1527$ and a {\Lya} absorber at $z = 0.2569$. From profile fitting, we find that the {\HI} component associated with the metal lines is fairly broad ($b \sim 50$~{\kms}). This component is undetected in {\Lyd}, as we expect it to be too weak. Our assumption of a simple three-component model for the {\HI} can be a source of systematic uncertainty in the line measurements. Given the low $S/N$ of data, we cannot ascertain whether the true kinematic nature of the {\HI} is more complex. Future observations involving the near-UV channel of COS will allow coverage for {\Lya} associated with this system, offering valuable constraints on the velocity profile of {\HI}. 

The {\OIV}~$788$, and {\NIV}~$765$ lines were also fitted with single component Voigt profiles. Figure \ref{fig4} shows the profile models for these various transitions. 

At the location of {\SV}~$786$ there is a very weak feature detected at $2.2\sigma$ significance.  We find this feature to be inconsistent with being {\SV} as none of the ionization models (discussed later) are able to simultaneously explain the detection of {\SV} with the non-detections of its adjacent ionization stages, namely {\SIV} and {\SVI}. It is quite likely that this is an unidentified interloping feature.  

\begin{figure}
\centering
\includegraphics[totalheight=0.35\textheight, trim=0cm 0cm -1cm 0cm, clip=true,angle=90]{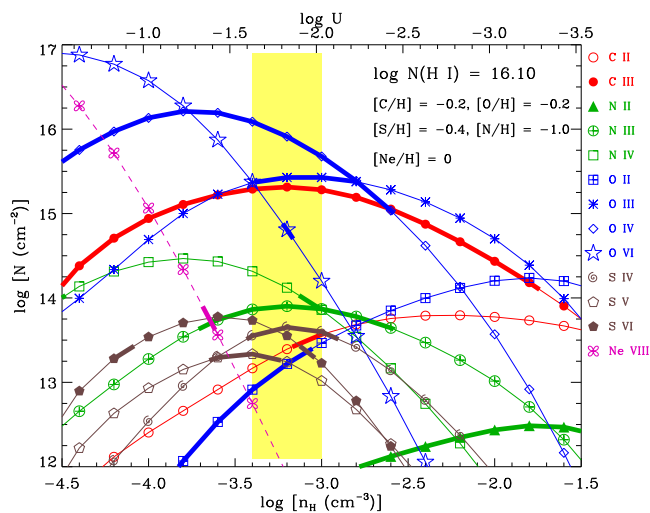}
\caption{Photoionization equilibrium models for the absorber at $z=0.61907$. The ionizing source is the KS15 extragalactic background radiation with $f_{esc} = 4\%$ escape fraction of hydrogen ionising photons. The thin curves show the model predicted column densities for the various ions at different densities. The thick portion in each curve is the region where the model is consistent with the observed column density to within $1\sigma$. The predicted column density ratios of $\dfrac{\NIII}{\NII}$, $\dfrac{\NIV}{\NIII}$, $\dfrac{\CIII}{\CII}$, $\dfrac{\SIV}{\SV}$ and $\dfrac{\SV}{\SVI}$ are simultaneously consistent with observations for the narrow density range of n$_{\H} \sim (0.3-1 \times 10^{3})$ {\cc}. This region (marked in \textit{yellow}) is the single phase solution for all the ions except {\NeVIII}.}
\label{fig6}
\end{figure}

\subsection{The {\NeVIIIdblt} Detection}

The {\NeVIIIdblt} absorption falls at a region of the spectrum where the $S/N \sim 12$ per wavelength bin, which is a factor of 2 higher compared to the $S/N$ at the {\Lyb} or the {\OVI} lines. The {\NeVIII}~$770$ feature has a rest-frame equivalent width of $W_r = 26~{\pm}~9$~m{\AA} when integrated over the velocity range from $-60$~{\kms} to $40$~{\kms}. This corresponds to a formal detection significance of $2.9 \sigma$. The statistical and the continuum placement errors have been taken into consideration in the error estimation. Including an additional $5$~m{\AA} to account for possible detector fixed pattern residual noise features would bring the detection significance down to $2.5 \sigma$. The redshifted wavelength of the accompanying {\NeVIII}~$780$ feature is contaminated by a strong {\Lyb} feature from an absorber at $z = 0.1944$ for which corresponding {\Lya}, {\CIII}, {\OVI} and several other metal lines are seen. The contamination renders a measurement on {\NeVIII}~$780$ impossible. 

In the individual exposures, the {\NeVIII}~$770$ has rest-frame equivalent widths of $W_r = 31~{\pm}~19$~m{\AA}, $W_r = 17~{\pm}~20$~m{\AA}, $W_r = 24~{\pm}~18$~m{\AA} and $W_r = 28~{\pm}~18$~m{\AA} for observations with G130M grating central wavelengths of $1291$~{\AA}, $1300$~{\AA}, $1309$~{\AA} and $1318$~{\AA} respectively. All of these observations have the same FP-SPLIT position. As with the previous case, we compared the spectra of a set of five quasars with the same COS grating settings and find no evidence for fixed pattern noise in this region of detector space. The absorption reported as {\NeVIII}~$770$ occurs at $\lambda = 1209.85$~{\AA}. This is not identified as an interloping line from other absorbers discovered along this sightline in the independent line identifications done by \cite{Danforth2016} and one of the co-authors (Bart Wakker). 

\section{Ionization \& Abundances in the $\lowercase{z}= 0.61907$ Absorber}\label{sec5}

\ To assess the density and temperature phases traced by this absorber, and the relative chemical abundances within them, we turn to time independent photoionization-recombination equilibrium models and collisional ionization models. We first describe the results from photoionization modelling. The models were computed using the standard photoionization package Cloudy (ver 13.03) last described by \cite{2013RMxAA..49..137F}. Cloudy models the absorbing gas as constant density plane parallel slabs irradiated by ionizing photons. The intensity and shape of the ionizing spectrum are assumed to be the extragalactic background radiation coming from AGNs and star-forming galaxies in the universe. The ionizing radiation field (hereafter EBR) we adopt is the upgraded  model of \citet[KS15]{KS2015b}, which incorporates the most recent measurements of quasar luminosity function from \citet{Croom2009} \& \citet{PD2013}, and star formation rate densities from \citet{KS2015a}. A source of uncertainty in the models for extragalactic background radiation is the escape fraction of Lyman continuum photons from star-forming galaxies. KS15 find that an escape fraction of $4\%$ is required to match the observed IGM {\HI} photoionization rate as measured by \cite{Kollmeier2014,Wakker2015} and 0\% to match the measurements by \citet{Shull2015}, \citet{Gaikwad2017a}, \citet{Gaikwad2017b}. Bearing in mind this uncertainty, we computed photoionization models for both $f_{esc} = 4$\% and $f_{esc} = 0$\%, and discuss the results for both cases. For brevity, we only display the modeling predictions for the $f_{esc} = 4$\% case. In the models, we assume the solar relative elemental abundances given by \cite{Asplund2009}. 

\subsection{Photoionization Equilibrium Models}

\ Figure \ref{fig6} shows the column density predictions from photoionization for the various ions at different gas densities. The models were generated for a {\HI} column density of $16.1$~dex, the value obtained from simultaneously fitting the Lyman series lines with a single component. The coverage of successive ionization stages of carbon, nitrogen, oxygen, and sulfur offer useful constraints for the ionization calculations. The density is best constrained by the measured column density ratios of $\log [N(\NIV)/N(\NIII)] = 0.27~{\pm}~0.16$, $\log [N(\SV)/N(\SIV)  = -0.24~{\pm}~0.06$, $\log [N(\SVI)/N(\SV) = 0.01~{\pm}~0.05$, and the lower limits $\log [N(\CIII)/N(\CII) \gtrsim 0.7$, $\log [N(\NIII)/N(\NII) \gtrsim 0.2$ and $\log [N(\OIII)/N(\OII) \gtrsim 1.9$. These ionic column density ratios are simultaneously valid for a gas density of $n_{\H} = (0.4 - 1) \times 10^{-3}$~{\cc}. At the mean value of $n_{\H} = 0.7 \times 10^{-3}$~{\cc}, the model predicts a total hydrogen column density of $N(\H) = 19.5$, a gas temperature and pressure of $T = 1.5 \times 10^4$~K and $p/K = 10.5$~{\cc}~K, and line of sight thickness of $L = 14.6$~kpc. The photoionization predicted temperature implies that the broadening of {\HI} and the intermediate and low ion lines are due to non-thermal motion. 

The single phase photoionization model with $f_{esc} = 4$\% suggests a near-solar abundance for C, and O. [C/H] is given by the unsaturated {\CII} column density. [C/H] is $\gtrsim -0.4$ for the model prediction to be consistent with the observed $N(\CII)$. Similarly, from the column density measurements of {\NIII}, and {\SIV} and the lower limit on {\OIII}, we obtain abundances of [N/H] $\gtrsim -1.0$, [O/H] $\gtrsim -0.2$ and [S/H] $\gtrsim -0.4$. For these ions to be coming from the same gas phase, the abundances have to be [C/H] = [O/H] = $-0.2$, [N/H] = $-1.0$, and [S/H] = $-0.4$. The single phase solution with these abundances is shown in Figure \ref{fig6}. 

\begin{figure}
\centering
\includegraphics[totalheight=0.36\textheight, trim=-1.2cm 0cm -1cm 0cm, clip=true]{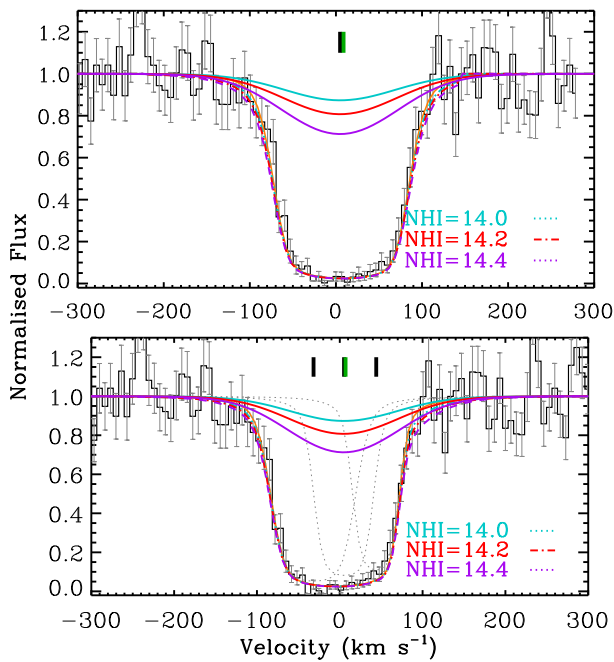}
\caption{The figure shows the {\HI} 1026 feature of the $z=0.61907$ absorber towards SDSS~J$080908.13+461925.6$. Superimposed on the data are different Voigt profile fits. The top and bottom panels assume a single component ($\chi^{2}_{\nu} = 1.7$) and a three component ($\chi^{2}_{\nu} = 1.2$) fit to the core absorption. The fit parameters are given in Table 1. To this are added broad {\HI} profiles with a fixed $b = 85$~{\kms} and $\log~N(\HI) = 14.0, 14.2, 14.4$~dex centered at the velocity where {\NeVIII} absorption is seen (show as solid curves of different color). The composite of the broad and narrow {\HI} absorptions are overlaid on the data as {\textit{dashed}} lines. The $b = 85$~{\kms} is how broad the {\HI} absorption from the {\NeVIII} gas phase would be, based on a temperature lower limit estimate for this phase. Addition of such a thermally broad and shallow {\HI} component affects only the wings of the profile and not the core absorption. From this, we estimate an upper bound to the {\HI} column density in the {\NeVIII} phase as $\log~N(\HI) \lesssim 14.2$~dex.}
\label{fig8}
\end{figure}

The predictions from the models with an EBR of $f_{esc} = 0$\% are not widely different. They yield similar density for the low ionization gas phase, with a $0.2$~dex increase in the relative elemental abundances. Also at energies $> 4$~Ryd, changing the spectral shape of the EBR within the measurement uncertainty of the AGN composite continuum given by \citet{stevans2014}, would result in a $\sim 0.2$~dex change in the hydrogen density. The abundance estimations thus carry an approximate uncertainty of ${\pm}~0.3$~dex because of the ambiguity in the escape fraction of ionizing photons, the spectral shape of the EBR, and from the uncertainty in the {\HI} column density. 

This single phase with $n_{\H} \sim 0.7 \times 10^{-3}$~{\cc} is also consistent with the observed $N(\OVI)$. For [O/H] = $-0.2$~dex, the photoionized gas phase simultaneously explains the observed {\OVI} along with its lower ionization stages. However, the predicted {\NeVIII} at this ionization parameter is $\sim 2$~dex smaller than the observed value for both versions of EBR with different $f_{esc}$. Producing the required amount of {\NeVIII} from the same gas phase would require increasing [Ne/H] by a factor of 100 from its solar value. The presence of {\NeVIII} thus points to a separate higher ionization phase in the absorber. This separate gas phase is unlikely to be dominantly photoionized for the following reasons. 

\begin{figure}
\centering
\includegraphics[totalheight=0.26\textheight, trim=0cm 0cm -1cm 0cm, clip=true]{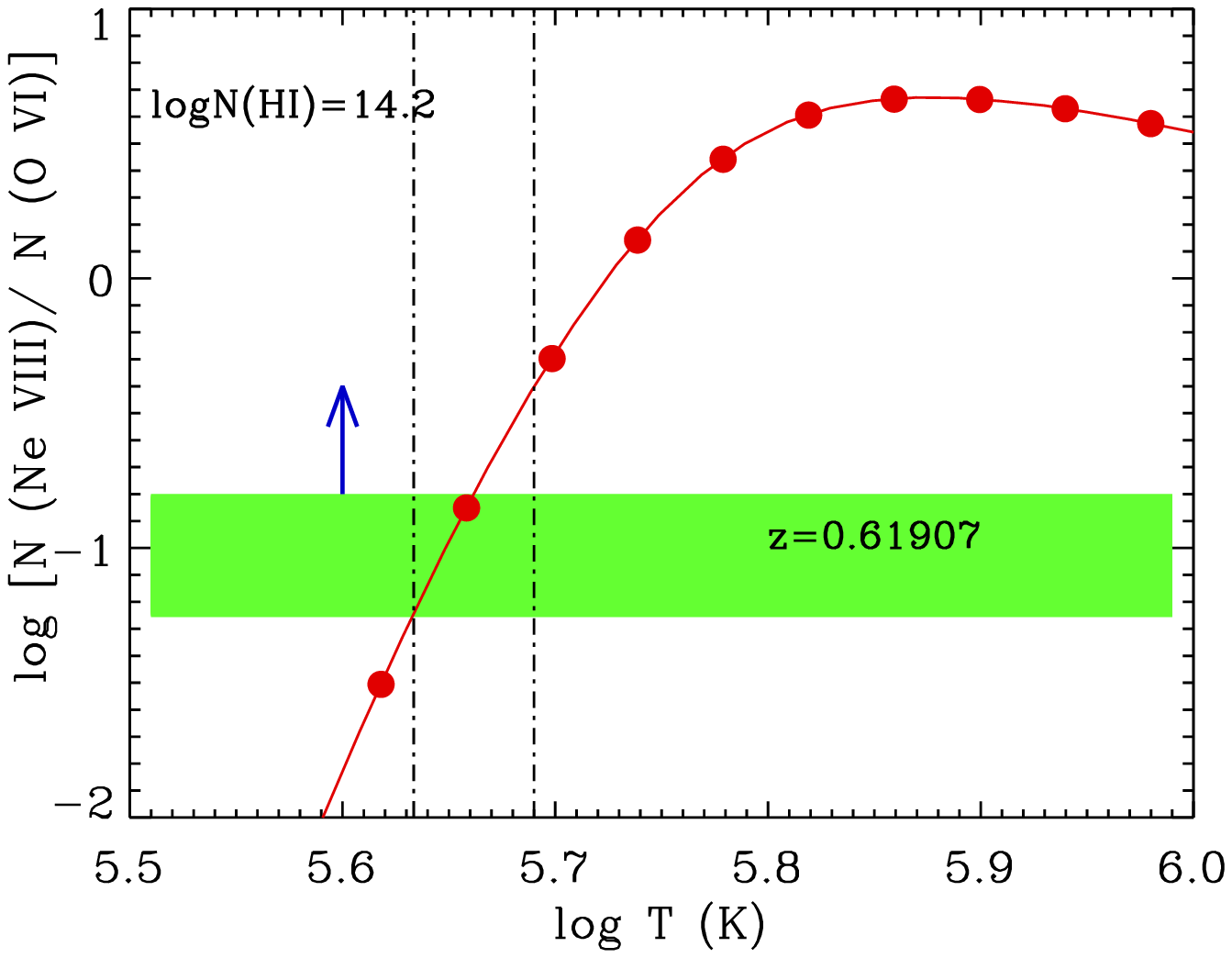}
\includegraphics[totalheight=0.26\textheight, trim=0cm 0cm -1cm 0cm, clip=true]{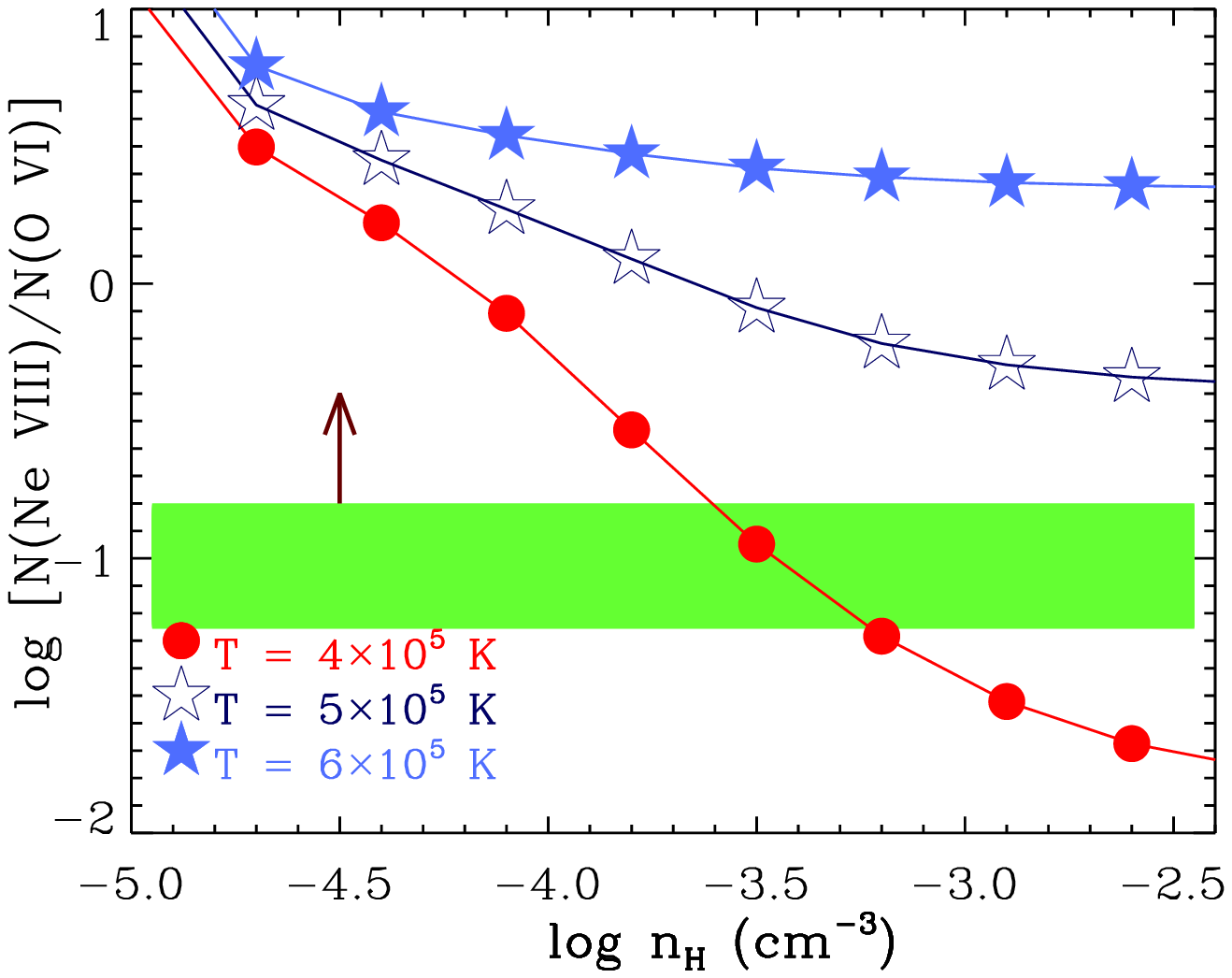}
\caption{ Figure on the \textit{top} panel shows the column density ratio of {\NeVIII}--{\OVI} predicted by CIE models of \citet{Gnat2007} for different equilibrium temperatures and solar elemental abundance. The green shaded region is where the model predicted ratio matches with the observation for the absorber at $z=0.61907$ towards SDSS~J$080908.13+461925.6$. The upward arrow is to indicate that the observed {\NeVIII} - {\OVI} column density ratio can be only considered as a lower limit on the true column density ratio between these two ions in the collisionaly ionized gas phase, since {\OVI} can have a significant contribution from the $T \sim 10^4$~K photoionized phase of this same absorber. The \textit{bottom} panel is the same ionic column density ratio predicted by photoionization - collisional ionization hybrid models for three different equilibrium temperatures of $T=4 \times10^{5}$ K, $T=5 \times10^{5}$K and $T=6 \times10^{5}$K respectively. The CIE models predict the {\NeVIII} to be from a warm gas phase with $T > 4.3 \times 10^5$~K. There is no clear constrain on the density from the hybrid models, since the {\NeVIII}--{\OVI} column density ratio is only a lower limit.}
\label{fig7}
\end{figure}

For solar [Ne/H], the observed $N(\NeVIII)$ is produced at ionization parameters of $\log U \geq -1.4$, corresponding to a densities of $n_{\H} \leq 2 \times 10^{-4}$~{\cc} and for $\log~N(\HI) \leq 16.1$. There are two discrepancies that emerge from such a separate higher photoionized phase. Firstly, at $\log U \sim -3.7$, this higher ionization phase also produces significant amount of {\CIII}, {\NIII}, {\OIII}, {\NIV}, {\OIV}, and {\OVI} even for low values of {\HI}. The combined column densities for these ions from the two gas phases will contradict the observed values by several factors. Secondly, the {\HI} column density associated with this high ionization gas is likely to be much smaller than the observed $\log~N(\HI) < 16.1$. To produce {\NeVIII} at lower values of $N(\HI)$ would require the line of sight to pass through very low-density columns of plasma that extend over several Mpc. For example, if the {\HI} associated with the {\NeVIII} gas phase has $\log N(\HI) \lesssim 14$~dex, then the observed $N(\NeVIII)$ will be recovered for solar [Ne/H] only for $\log U \gtrsim -0.5$ corresponding to $\log n_{\H} \lesssim -4.5$. The absorbing region, in this case, has to be spread over a large path length of $\gtrsim 0.5$~Mpc, which is nearly equal to the full virial cross-section of $L^*$ galaxies ($R_{vir} \sim 200 - 300$~kpc). It is unlikely for gas spread over such a large length to maintain velocity dispersions of a few tens of {\kms}. The other possibility is that the absorber mass has not yet decoupled from the universal expansion. In such a case, the absorption lines will suffer a velocity broadening due to Hubble expansion, which will be $v(z) = H(z)~L \sim 50$~{\kms}, where $L \geq 0.5$~Mpc. The observed $b$-value for the well measured {\HI} lines are $40$\% narrower than the expected broadening due to Hubble flow. 

We emphasize here that the photoionization models are not exact because of the simplistic assumptions built into them (the absorbing cloud in the models have a plane parallel geometry with uniform density and temperature) and also due to the lack of information on the exact column densities and any sub-component structure in the saturated metal lines and {\HI}. However, the model prediction that the {\NeVIII} is not consistent with photoionization is important. In the next section, we discuss the results from collisional ionization models for the {\NeVIII} bearing gas.

\subsection{Evidence for a Warm Gas Phase}

\subsubsection{Collisional Ionization Equilibrium Models}

In collisional ionization equilibrium (CIE) models, the ionization fractions of elements depend only on the equilibrium temperature. In Figure \ref{fig7} we plot the {\NeVIII} to {\OVI} column density ratio predictions made by the CIE model of \cite{Gnat2007} for a range of plasma temperatures. The CIE is a good approximation for calculating the ionization of {\HI}, {\NeVIII} and {\OVI} at $T > 2 \times 10^5$~K since the gas cools relatively slowly at such higher temperatures. For lower temperatures, non-equilibrium ionization effects (recombination lagging behind the cooling of the gas) become important \citep{Savage2014}.

The column density ratio of {\NeVIII} to {\OVI} cannot be used directly to ascertain the temperature of the gas, as much of the {\OVI} is possibly tracing the cooler photoionized medium. Nonetheless, the presence of {\OVI} can be used to place a lower limit on the temperature of the collisionally ionized gas. As shown in the bottom panel of Figure \ref{fig7}, the ratio of $\log~[N(\NeVIII)/N(\OVI)] \gtrsim -1.25$ is valid for $T \gtrsim 4.3 \times 10^5$~K. If this warm gas phase has more {\NeVIII} compared to {\OVI}, then a lower limit of $T \gtrsim 5.0 \times 10^5$~K  is obtained from the CIE models. 
\begin{figure*}
\centering
\includegraphics[totalheight=0.35\textheight, trim=0cm 0cm -1cm 0cm,clip=true,angle=90]{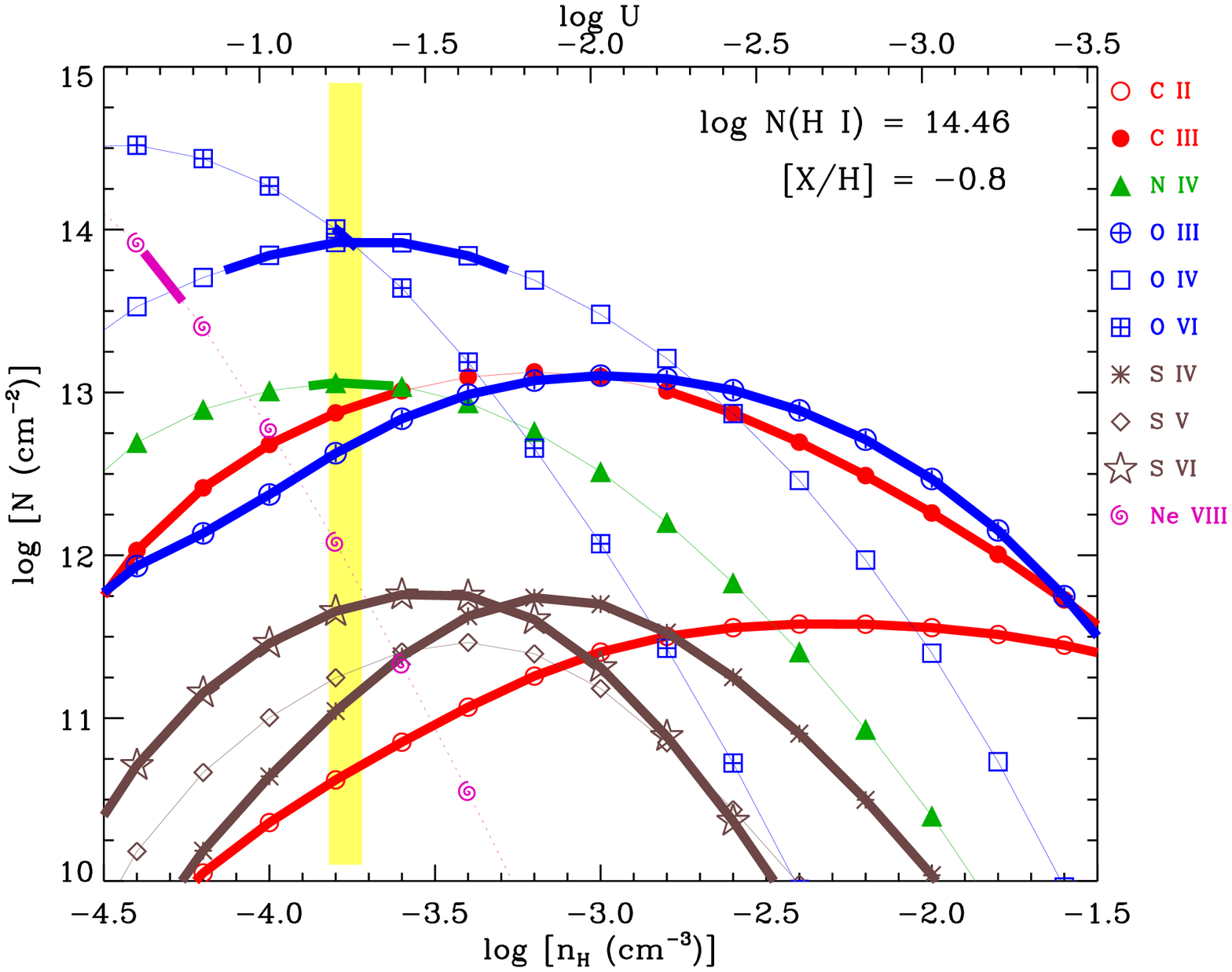}
\includegraphics[totalheight=0.35\textheight, trim=0cm 0cm -1cm 0cm,clip=true,angle=90]{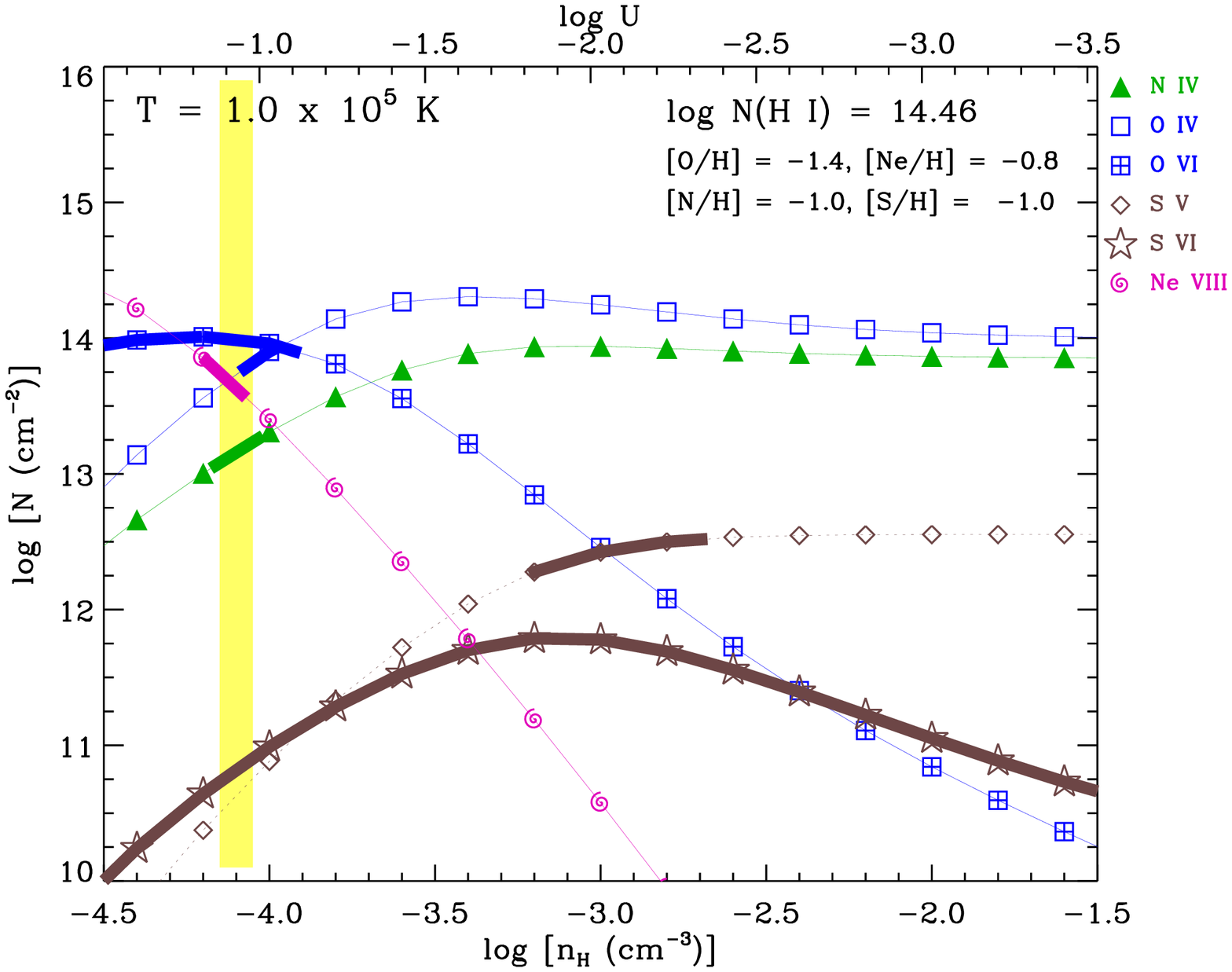}
\caption{The figure on the left panel is the photoionization equilibrium (PIE) model for the absorber at $z=0.57052$ with a $f_{esc} = 4\%$ in the KS15 ionizing background. The curves correspond to column density predictions at different densities for each ion, and the thick regions on each curve is where the model is consistent with the observation to within $1\sigma$. For a sub solar metallicity of $-0.8$ dex and at a density of $n_{\H} \sim 5 \times 10^{-3}$\cc the PIE model can explain the observed column density of all the ions except {\SV} and {\NeVIII}. The right panel shows the hybrid ionization model predictions for a temperature T=$1 \times 10^{5}$~K and the KS15 EBR with $f_{esc} = 4\%$. The temperature used is based on comparing $b(\OVI)$ with $b(\HI)$. In the hybrid models, the various ions are produced from a single phase with $n_{\H} \sim 10^{-4}$~{\cc} for elemental abundances of [O/H] $= -1.4$, [Ne/H] $= -0.8$, [N/H] $= -1.0$ and [S/H] $= -1.0$.}
\label{fig9}
\end{figure*}
To determine the metallicity and the total hydrogen column density in this warm phase, we need an estimate on the associated {\HI}. Even at the conservative lower limit of $T = 4.3 \times 10^5$~K, the neutral fraction of hydrogen at CIE is $f(\HI) = 7.9 \times 10^{-7}$, suggesting that most of the mass in the {\NeVIII} gas phase is in an ionized form. Consequently, nearly all of the strong absorption in {\HI} is potentially tracing the cooler photoionized medium. Given the low {\HI} optical depth and the warm temperature of the gas, we expect the absorption from the residual {\HI} to be thermally broad and shallow. The absence of coverage of {\Lya} makes it implausible to know whether a BLA is associated with the {\NeVIII} gas. At the low $S/N$ of the data, it is difficult to search for the presence of a broad {\HI} feature in the weaker {\Lyb} line. We, therefore, use the {\Lyb} to place a useful upper limit on the {\HI} column density associated with this warm gas. 

In Figure \ref{fig8} are superimposed synthetic BLA profiles on top of the {\Lyb} absorption. The BLA profiles were synthesized for the CIE temperature lower limit of $T = 4.3 \times 10^5$~K with different {\HI} column densities. The temperature corresponds to a pure thermal line width of $b(\HI) = 85$~{\kms}. We explored both single component and three component (refer Table \ref{tab1}) models for the narrow and strong absorption that forms the core of the {\HI} profile. The three component model is based on the weakly resolved kinematic sub-structure seen in the {\HI}~$923$ and $926$ lines, and also {\NIV}~$765$. Statistically, the three components result in a better fit ($\chi^{2}_{\nu} = 1.2$) to the core absorption in {\Lyb} compared to a single component ($\chi^{2}_{\nu} = 1.7$), although in the latter case the fitting model is within $1\sigma$ of the flux values. Irrespective of whether the core absorption is modeled by a single component or multiple components, the thermally broad {\HI} associated with the {\NeVIII} gas phase has to have $\log~N(\HI) \lesssim 14.2$ to go undetected in the {\Lyb} profile. If the temperature in the {\NeVIII} gas phase is higher, this {\HI} limiting column density will also be higher. The {\HI} column density upper limit of $\log N(\HI) = 14.2$ constrains the [Ne/H] $\lesssim -0.5$ in the warm gas. 

Adopting $\log~[N(\HI)] \lesssim 14.0$, CIE models predict a total hydrogen column density of $\log~[N(\H)] \gtrsim 20.1$ for $T \gtrsim 4.3 \times 10^5$~K. This lower bound on baryonic column density is an order of magnitude more than the amount of total hydrogen present in the photoionized gas phase. 

Extending the collisional ionization scenario further, we also computed models that simultaneously consider ion-electron collisions and ion-photon interactions as sources of ionization. Insights from these \textit{hybrid} models are discussed next.  

\subsubsection{Hybrid of Photoionization \& Collisional Ionization Equilibrium}

Using Cloudy we computed \textit{hybrid} models that simultaneously consider photoionization and collisional ionization equilibrium scenarios for different warm gas temperatures. The photoionizations are caused by the extragalactic ionizing background radiation as given by KS15. The {\NeVIII} to {\OVI} column density ratio predicts a temperature lower limit of $T \sim 4 \times 10^5$~K from CIE. At this temperature, the hybrid models require gas densities of $n_{\H} < 10^{-3}$~{\cc} to be consistent with the lower bound of the {\NeVIII} to {\OVI} column density ratio. The number density corresponds to a baryonic overdensity of $\Delta = \rho/\bar{\rho} \gtrsim 1000$ (using $n_{\H} (\cc) = 1.9 \times 10^{-7}~(1+z)^3~\Delta$). Simulations associate such $T - \Delta$ combinations typically with hot halos (Figure 5 of \citet{Gaikwad2017a}). For the approximate {\HI} upper limit of $N(\HI) = 10^{14.2}$~{\cmsq} derived earlier, this gas phase model predicts, $N(\H) \sim 4.3 \times 10^{20}$ ~{\cmsq} and an absorption path length of $L \sim 88$~kpc. Such a prediction is also consistent with the lower limit on {\NeVIII} to {\SVI}. At that temperature and {\HI} column density, the [Ne/H] = $-0.5$~dex to match the observed $N(\NeVIII)$ at any density. Higher temperatures for the {\NeVIII} gas phase would predict higher density upper limits (as shown in Figure \ref{fig7}). We emphasize here that given the lack of observational constrain on the amount of {\OVI} and {\HI} associated with the {\NeVIII}, the hybrid models are only as useful as the CIE models in setting limits on the physical conditions of the warm gas. Furthermore, if the true temperature of the gas is $T \lesssim 2 \times 10^5$~K, the hybrid models should assume non-CIE calculations along with photoionization for a truly valid explanation of the physical conditions. Such involved modeling is beyond the scope of this paper (but, see \citet{Oppenheimer2016} where this is attempted).  

\section{Ionization \& Abundances in the $\lowercase{z}=0.57052$ Absorber towards SBS~$1122+594$}\label{sec6}

The absence of low ionization species like {\CII}, {\OII}, {\CIII} and {\OIII} suggests moderate to high ionization conditions in this absorber. Photoionization predicted column densities using Cloudy for the various metal ions are shown in Figure \ref{fig9}. The models were computed for an {\HI} column density of $14.46$~dex which we measure for the $v \sim -8$~{\kms} component. The metal ions are all coincident in velocity with this {\HI} component. The ratio between the observed {\OIV} and {\OVI} is valid for a density of $n_{\H} = 1.8 \times 10^{-4}$~{\cc}. At this density, the models are able to recover the observed column density of {\NIV}, {\OIV}, {\OVI} for a metallicity of [X/H] $= -0.8$~dex. This single phase solution, shown in Figure \ref{fig9}, is also consistent with the non-detection of the low ionization species. The single phase model with $n_{\H} = 1.8 \times 10^{-4}$~{\cc} yields a total hydrogen column density of $N(\H) = 18.6$, a temperature of $T = 2.7 \times 10^4$~K, a gas pressure of $p/k = 4.9$~{\cc}~K and a line of thickness of $L = 47.2$~kpc for the absorbing medium. 

The {\NeVIII} column density from this gas phase is $\sim 1.8$~dex lower than the observed value. To explain {\NeVIII} from the same phase would require the [Ne/H] to be $\sim 10$ greater than solar, whereas the C, N, O abundances are significantly sub-solar. The alternative of the {\NeVIII} arising in a separate photoionized phase at higher ionization parameter can also be ruled out, as such a phase will also produce significant {\OVI}, and {\OIV} in it, which will make the two-phase solution incompatible with the observed column densities of these ions. 

There is also a contradiction in the temperature arrived at through photoionization modelling and the measured $b$-values of {\HI} and {\OVI}. The different b-values for the reasonably well aligned {\HI} and {\OVI} suggests the temperature of this gas phase to be $T = (0.5 - 1.5) \times 10^5$~K, where the range corresponds to the $1\sigma$ uncertainty in the $b$-values. Similar limits for temperature are obtained if we use the b-values of {\HI} and {\OIV}. The values are too high for UV photoionization heating. Such a conclusion rests on the assumption that the metal lines and {\Lya} are adequately resolved by COS. In the {\OIV}~$788$, {\SV}~$786$ and {\OVI}~$1032$ lines there is no immediate evidence for kinematic substructure. The {\NIV}~$765$ shows slight assymetry in the bluer side of its  profile. At the given resolution and $S/N$, it is difficult to rule out whether the lines are narrower than what is seen by COS. 

The low $S/N$ of the data not withstanding, the higher temperatures suggested by the line widths could mean that the collisional processes may be playing a significant role in controlling the ionization in this gas. A more realistic model for this \textit{warm} plasma could be one where electron collisions as well as photon interactions are simultaneously considered. Such \textit{hybrid} models are discussed in the next section. 
\begin{figure*}
\centering
\includegraphics[width=0.48\textwidth, trim=0cm 0cm 0cm 0cm, clip=true]{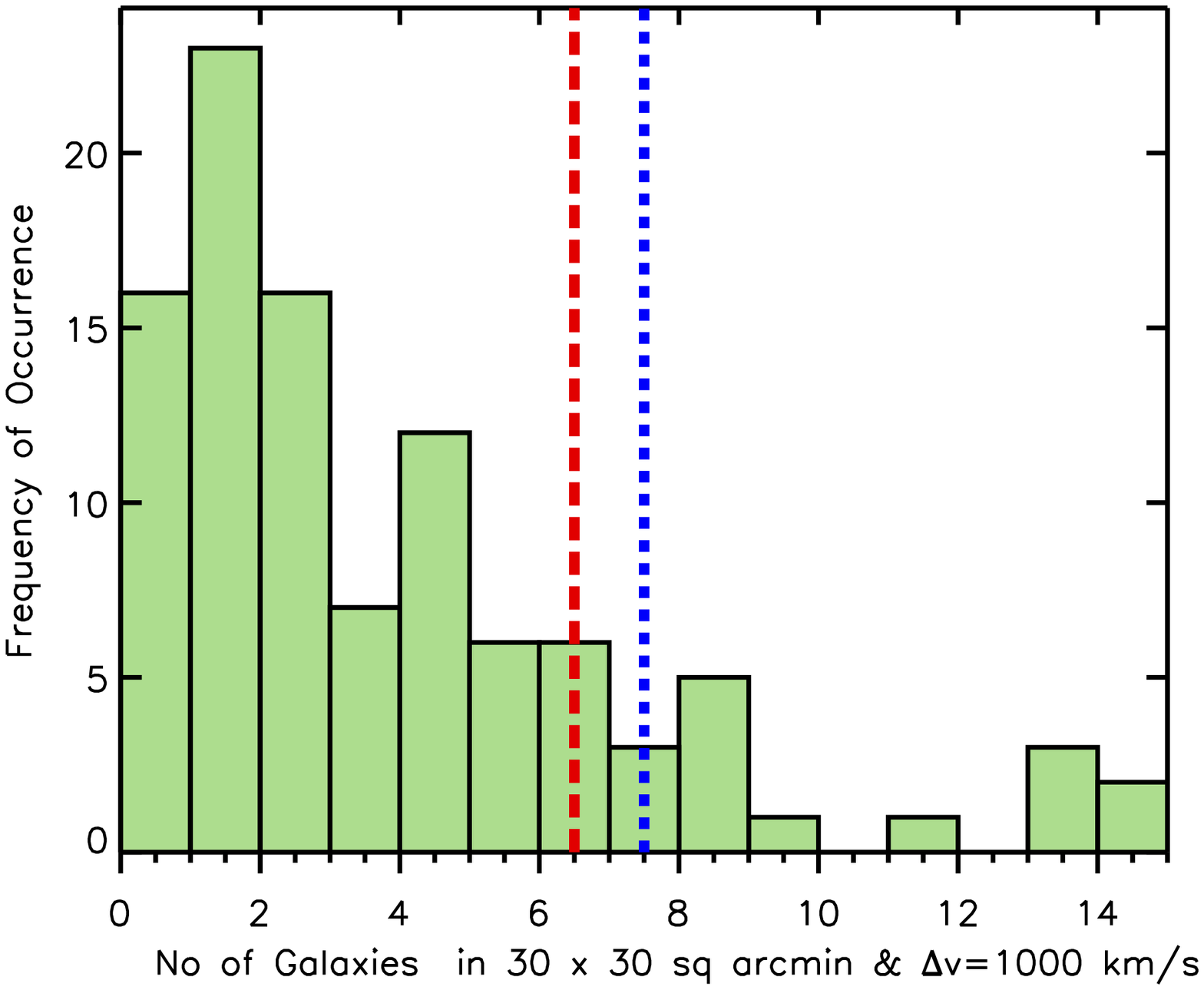}
\includegraphics[width=0.46\textwidth, trim=0cm 0cm 0cm 0cm, clip=true]{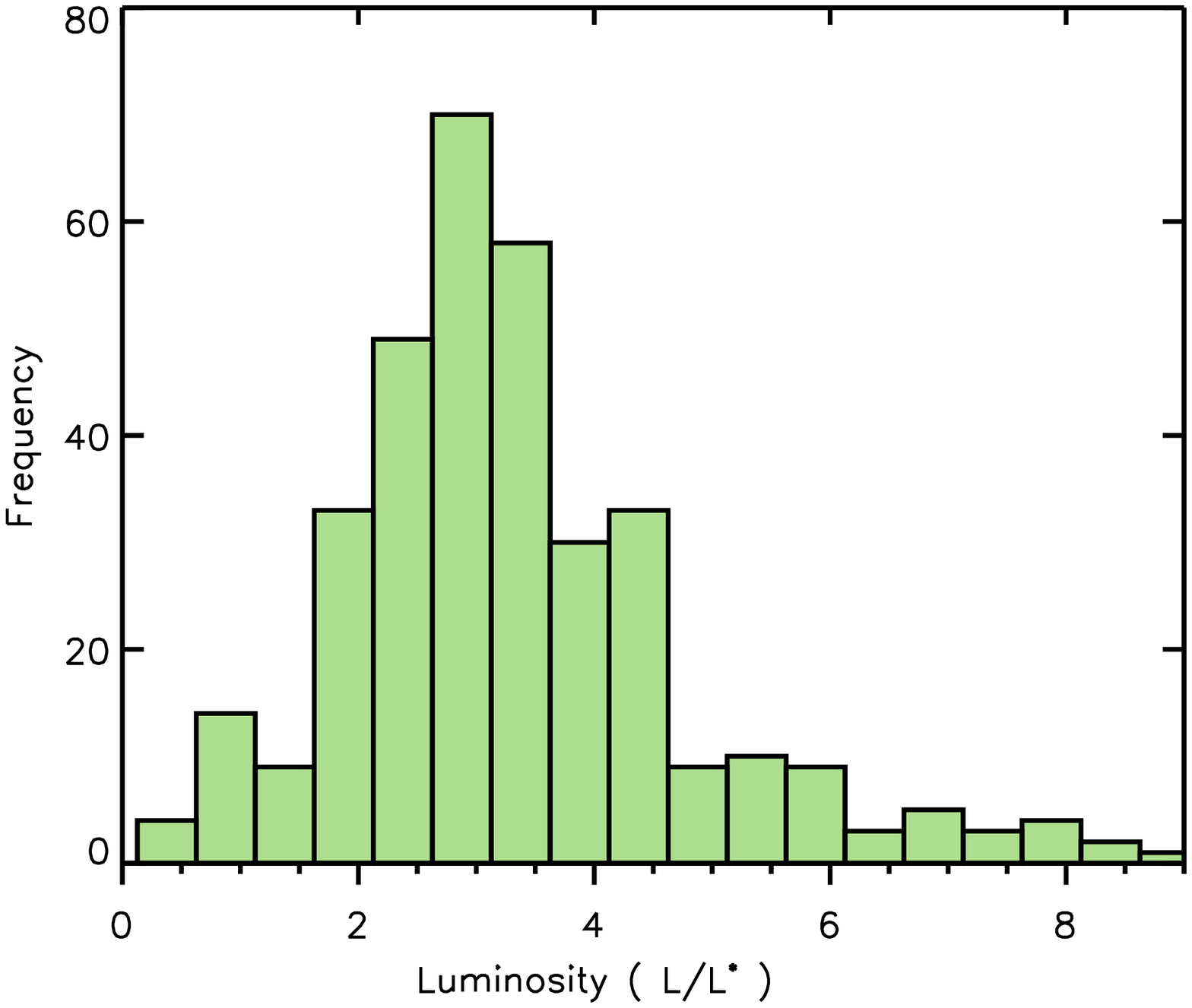}
\caption{The \textit{left panel} shows the distribution of galaxies within $30 \times 30$~arcmin and $|\Delta v| = 1000$~{\kms} of $z=0.6$ in 100 locations randomly sampled from the SDSS footprint. The distribution indicates that there is only a small ($15$\%) probability of finding more than 6 galaxies in the sampled volume. More than half the number of times ($\sim 56$\%) one finds only 2 galaxies or less in the chosen search window, implying that the galaxy overdensity seen near to both {\NeVIII} absorbers is not random coincidence. The \textit{dashed} and \textit{dotted} vertical lines indicate the number of galaxies found in the same region of space around the $z = 0.57052$ and $0.61907$ {\NeVIII} absorbers. In the \textit{right panel} is the luminosity distribution of the galaxies from the random sampling, indicating that the completeness of SDSS is poor for $L \lesssim 3L^*$ at $z = 0.6$, which is inturn consistent with the large luminosities that we find for the galaxies in the extended environment around both {\NeVIII} absorbers.}
\label{no_gal}
\end{figure*}
\subsubsection{Hybrid of Photoionization \& Collisional Ionization Equilibrium Models}

In Figure \ref{fig9}, we have also shown the predictions from hybrid models for an EBR with $f_{esc} = 4$\%. The temperature of the plasma was set to $T = 10^5$~K, which is the mean of the temperature range obtained from the thermal broadening of {\HI} and metal absorption lines. This temperature is closer to where {\OVI} reaches its peak ionization fraction ($T = 3 \times 10^5$~K), compared to the peak in {\NeVIII} ($T = 7 \times 10^5$~K). From the observed {\OIV} to {\OVI} ratio, the density is constrained to $n_{\H} \sim 10^{-4}$~{\cc}, with only a slight difference between the predictions from the two flavors of EBR. 

The hybrid models suggest that it is possible for {\NIV}, {\OIV}, {\OVI} and {\NeVIII} to be coming from the same gas phase with $n_{\H} \sim 10^{-4}$~{\cc}, for non-solar relative elemental abundances. This gas phase model yields a total hydrogen column density of $N(\H) = 4.8 \times 10^{19}$~{\cmsq}, a line of sight thickness of $L \sim 196$~kpc, and represents a baryonic overdensity of $\Delta \sim 100$. The observed column densities are recovered for [O/H] $= -1.4$, [Ne/H] $= -0.8$, and [N/H] $= -1.0$. The baryonic column density and absorber size reduces to half, and the elemental abundances increase by $0.2$~dex for hybrid models with $f_{esc} = 0$\%. 

The higher than solar (Ne/O) ratio may appear unusual. However, the solar abundance of Ne is a poorly determined quantity. The lack of strong photospheric Ne transitions in the optical or UV is the major source of uncertainty in solar Ne abundance measurements. From X-ray spectroscopic observations of a sample of stars in the 100 pc neighborhood of the Sun, \cite{Drake2005} estimate a value for (Ne/O) that is $2.3$ times higher than the solar value of (Ne/O) $= -0.76~{\pm}~0.11$ given by \cite{Asplund2009}. Adopting this revised estimate for (Ne/O) with an increase of $+0.3$~dex will make the [Ne/H] $\sim$ [O/H] in the warm phase of the absorber.  
\begin{figure*}
\centering
\includegraphics[totalheight=0.4\textheight, trim=0cm 0cm -1cm 0cm, clip=true]{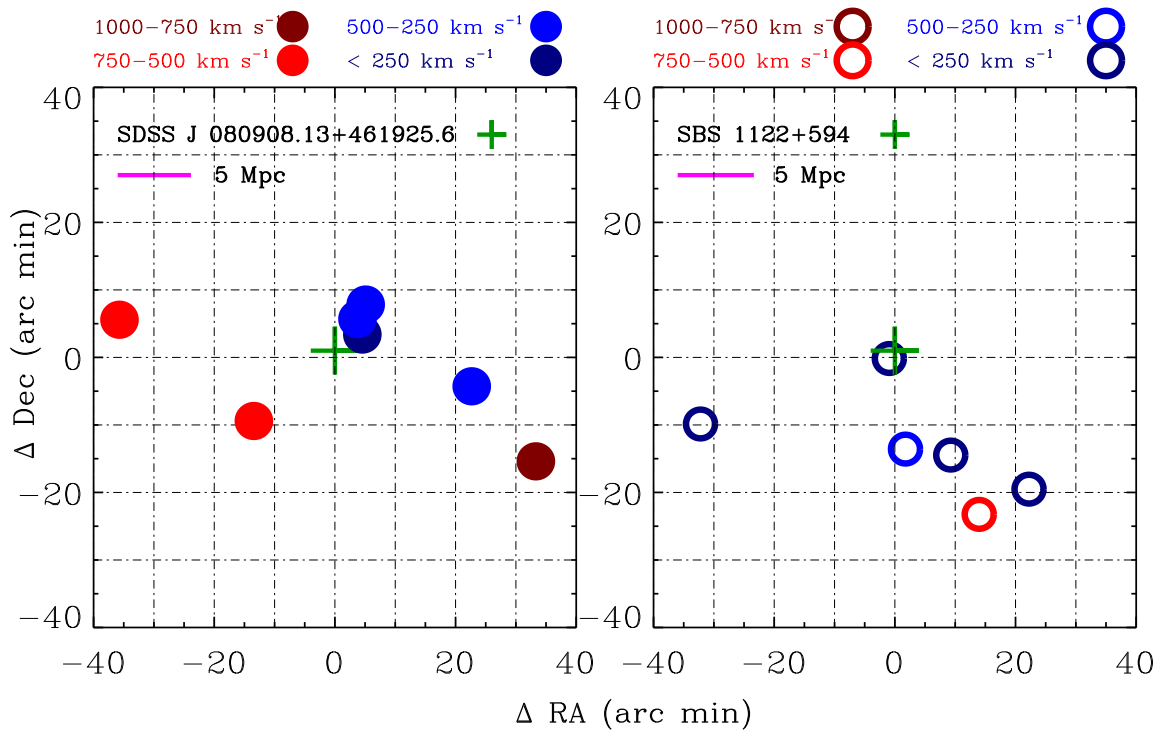}
\caption{Figure shows the galaxies observed in SDSS survey within $30 \times 30 $ arcmin and 1000~{\kms} from the absorbers detected at $z=0.61907$ (left panel) and $z=0.57052$ (right panel) towards the QSO sightlines SDSS $J080908.13+461925.6$ and SBS~$1122+594$ respectively. Galaxies associated with $z=0.61907$ absorber is shown with filled circles and those associated with $z=0.57052$ is shown with open circles where the colour coding corresponds to the velocity separation of the galaxy from the absorber. The line of sight is indicated by the "+" sign.}
\label{fig11}
\end{figure*}


\section{Galaxies Near the Absorber}\label{sec7}

In this section, we discuss the galaxies that lie in the neighbourhood of the absorbers. Both sightlines are covered by the SDSS survey. The SDSS DR12 spectroscopic database is $90$\% complete down to an $r$-band magnitude of $r < 17.8$ \citep{Strauss2002}. This translates into a luminosity of $\gtrsim 6~L*$ at $z \sim 0.6$ \citep{Ilbert2005}, implying that the SDSS is sampling only the very brightest galaxies at the redshifts of these absorbers.     
\begin{table*}
\centering
\caption{Galaxies associated with the absorbers} 
\scriptsize
\begin{tabular}{l l l r c c c c c c c}
\hline
R.A.  &  Dec.  &  $z$ & $\Delta v$ {\kms}   &  $\eta$ (arcmin)   & $\rho$ (Mpc)  & $g$(mag)& $r$(mag) & M$_g$ &($L/L^*$)$_g$ \\[2ex] 
\hline
\hline
\multicolumn{9}{c}{$z_{abs} = 0.61907$ ($v = 0$~{\kms}) Absorber towards SDSS $J080908.13+461925.6$} \\ \hline \hline
$122.360  $ & 46.3802 & $0.6199 \pm 0.0002$  & $153 \pm 40$ & 4.62868 & 1.9 &  $ 24.4 \pm  0.9  $ &$ 21.0 \pm 0.1$ &  -18.9 & 0.1 \\
$122.346  $ & 46.4186 & $0.6207 \pm 0.0002$  & $301 \pm 40$ & 6.25086 & 2.6  &  $ 22.0 \pm  0.1  $ &$ 21.2 \pm 0.1$ &  -22.0 & 2.3 \\
$122.662  $ & 46.2525 & $0.6212 \pm 0.0002$  & $385 \pm 40$ & 16.2716 & 6.7  &  $ 23.7 \pm  0.7  $ &$ 21.7 \pm 0.2$ &  -21.8 & 1.9 \\
$122.060  $ & 46.1669 & $0.6223 \pm 0.0002$  & $590 \pm 40$ & 13.2152 & 5.5  &  $ 23.0 \pm  0.4  $ &$ 20.7 \pm 0.1$ &  -21.4 & 3.6 \\
$122.369  $ & 46.4542 & $0.6215 \pm 0.0003$  & $449 \pm 55$ & 8.58469 & 3.5  &  $ 22.7 \pm  0.3  $ &$ 20.9 \pm 0.2$ &  -22.7 & 4.3 \\
$121.689  $ & 46.4168 & $0.6157 \pm 0.0001$  & $-620 \pm 30$ & 25.2369 & 10.4 &  $ 20.7 \pm  0.1  $ &$ 20.1 \pm 0.1$ &  -22.8 & 5.1 \\
$122.839  $ & 46.0672 & $0.6239 \pm 0.0002$  & $897 \pm 40$ & 27.7513 & 11.4 &  $ 23.4 \pm  0.4  $ &$ 21.7 \pm 0.2$ &  -21.9 & 2.1 \\       
$122.746  $ & 46.4922 & $0.61   \pm 0.04^p$  &              & 21.6522 & 8.9  &  $ 23.0 \pm  0.2  $ &$ 21.2 \pm 0.1$ &  -22.3 & 3.3 \\
\hline
\multicolumn{9}{c}{$z_{abs} = 0.57052$ ($v = 0$~{\kms}) Absorber towards SBS~$1122+594$} \\ \hline \hline
$171.45883 $ &  59.169975 & $0.570542 \pm 0.00014$ & $   8 \pm  30 $  &  0.497249 & 0.197 & $22.3 \pm 0.2 $ & 20.9 $\pm$ 0.1 & -22.2 & 2.6\\      
$171.50348 $ &  58.946364 & $0.568724 \pm 0.00017$ & $-339 \pm  50 $  &   13.613460 & 5.395 & $22.8 \pm 0.3 $ & 21.0 $\pm$ 0.1 & -22.1 & 3.0 \\ 
$171.62853 $ &  58.931547 & $0.569372 \pm 0.00050$ & $-215 \pm 100 $  &   15.236410 & 6.039 & $22.2 \pm 0.2 $ & 20.6 $\pm$ 0.1 & -22.6 & 4.0 \\ 
$170.93717 $ &  59.008323 & $0.570136 \pm 0.00020$ & $ -69 \pm  70 $  &   19.265253 & 7.635 & $22.1 \pm 0.1 $ & 20.8 $\pm$ 0.1 & -22.3 & 3.2 \\ 
$171.70715 $ &  58.785227 & $0.573197 \pm 0.00019$ & $ 514 \pm  70 $  &   24.341730 & 9.648 & $21.2 \pm 0.1 $ & 20.1 $\pm$ 0.1 & -22.9 & 5.4 \\ 
$171.84478 $ &  58.847470 & $0.570971 \pm 0.00018$ & $  90 \pm  50 $  &   22.627742 & 8.968 & $22.3 \pm 0.2 $ & 20.7 $\pm$ 0.1 & -22.4 & 3.4 \\
\hline
\hline
\end{tabular}
\label{tab3}
\begin{flushleft}
{ Comments: Information on the galaxies identified by SDSS within ${\pm}$1000 {\kms} and $30 \times 30$ arcmin$^2$ from the two absorbers are listed here. The equatorial coordinates of the galaxies are in the first two columns. The third column lists the spectroscopic redshifts of the galaxies as given by SDSS. The line of sight velocity separation between the absorber and the galaxy is in the fourth column. The angular separation between the absorber and the galaxy in the plane of the sky, and the corresponding projected physical separation are given in the next two columns. The projected separation was estimated assuming a $\Lambda$CDM universe with $H_0 = 69.6$~{\kms}~Mpc$^{-1}$ , $\Omega_{m} = 0.286$ and $\Omega_{\Lambda} = 0.714$. The apparent magnitude in the $g$-band is given in column 7. The absolute magnitude listed in column 8 was K-corrected using the emperical K-correction given by \cite{Westra2010}. The last column provides the luminosity in terms of the Schechter luminosity of $M_{\star}=-21.06$ at $z=0.6$ obtained from \citet{Ilbert2005}. For estimating the projected separation and luminosity distance to these galaxies, we have used the cosmology calculator of \citet{Wright2006}. One of the galaxies (labeled $p$) has only photometric redshift available.}
\end{flushleft}
\end{table*}

\ Within a projected separation of $ 30 \times 30 $ arcmin from the line of sight towards SDSS $J080908.13+461925.6$, there are 7 galaxies with spectroscopic redshifts that place them within $|\Delta v| = 1000$~{\kms} rest-frame velocity of the $z=0.61907$ absorber. Information on the galaxies are listed in Table \ref{tab3}, and their relative locations are shown in Figure \ref{fig11}. The impact parameters of these galaxies range from $0.7 - 11.4$~Mpc. The galaxy nearest to the absorber, though close-by in velocity ($|\Delta v| = 152.9$~{\kms}), is at a projected separation of $1.91$~Mpc. The halo radius of this galaxy can be estimated from its scaling relationship with luminosity given by \cite{stocke2014}. 

\begin{equation}
    \log R_{vir} = 2.257 + 0.318\mathrm{C} + 0.018 \mathrm{C^2} - 0.005 \mathrm{C^3}
\end{equation}

where $\mathrm{C} = \log (L/L^*)$. The estimated luminosity of $0.1L^*$ suggests a halo radius as $R_{vir} = 91$~kpc, which is 20 times smaller than the projected separation in the plane of the sky between the absorber and the galaxy. Given this large separation, the {\NeVIII} - {\OVI} absorber is unlikely to be coming from gas embedded within the hot halo of the galaxy.  We cannot rule out the possibility of the absorber being associated with a $\leq L^*$ galaxy (or even brighter) closer to the sightline, but undetected by SDSS. Indeed, galaxies close-by in velocity and physical separation to {\NeVIII} absorbers are known to span a wide range in luminosities from sub-$L^*$ to $> L^*$ \citep{Chen2009,Mulchaey2009,Tripp2011,Meiring2013}. 

The 7 galaxies that are coincident in redshift with the absorber have a narrow velocity dispersion of $\sigma \sim 85$~{\kms} and an average systemic velocity that is $\sim 485 \pm 16$~{\kms} with reference to the absorber redshift. The abundance of $\gtrsim L^*$ galaxies within such a narrow range of projected physical separation and velocity offset indicates that the line of sight is possibly passing through a group medium. Given the limited number count of galaxies at the redshift of the absorber and the incompleteness of the sample for even $L*$ luminosities, it is not realistic to formally define a galaxy group through a friends of friends approach, or similar standard algorithms. 

We used random sampling to investigate whether the line of sight is indeed probing an overdensity region, or if the detection of galaxies is consistent with their random distribution in space. We selected 100 different locations within the SDSS footprint at random and searched for galaxies at $z = 0.6$ that lie within $30 \times 30$~arcmin and $|\Delta v| = 1000$~{\kms} of each location. The frequency distribution from this random sampling, shown in Figure \ref{no_gal}, suggests that there is only a $15$\% chance of finding more than 6 galaxies within the sampled volume. During more than half the number of times ($\sim 56$\%), our sampling found only 2 galaxies or less in the search window, indicating that the region intercepted by SDSS $J080908.13+461925.6$ at the location of the absorber is most likely an overdensity region, such as a group or cluster environment. 

The $z = 0.57052$ absorber towards SBS~$1122+594$ also resides in a galaxy overdensity region with $6$ galaxies identified within $|\Delta v| = 1000$~{\kms} of the absorber. The distribution of galaxies is shown in Figure \ref{fig11} with open circles, and their information is listed in Table \ref{tab3}. The nearest galaxy with a spectroscopically confirmed redshift is nearly at the systemic velocity of the absorber, at a close-by projected separation of $\rho = 197$~kpc. The $L = 2.6~L^*$ luminosity of the galaxy suggests that the absorber is residing well within the galaxy's virial radius of $R_{vir} = 245$~kpc. The impact parameter is also comparable to the size of {\OVI} absorbing halos around luminous galaxies at low-$z$ \citep{Tumlinson2011,Werk2013,Muzahid2014}. The second closest galaxy to the absorber is at a projected distance of $\sim 5$~Mpc and displaced in velocity by $|\Delta v| \sim 339$~{\kms}. The average systemic velocity of the galaxies relative to the absorber is $-27 \pm 26 $~{\kms} and the velocity dispersion is $ 111$~{\kms}. 

The densities predicted by the hybrid models correspond to overdensities ($\Delta = \rho/\bar{\rho} \gtrsim 10^2$) that are reminiscent of hot halos and the warm-hot intergalactic gas in simulations \citep{Smith2011,Gaikwad2017a}. The $z = 0.57052$ absorber is well within the virial radius of a $2.6~L^*$ galaxy. Using the $r$-band magnitude and halo mass ($M_h$) relationship given by \cite{Tinker2009}, we estimate that this galaxy resides in a halo of minimum mass $M_h \sim 10^{14}$~M$_{\odot}$. The corresponding virial temperature of the dark matter halo comes out to be $T_{vir} \sim 10^7$~K. Since the halo mass is greater than the critical mass of $M_h \sim 10^{11}$~M$_\odot$ for virial shocks \citep{Birnboim2003}, any gas accreted by the galaxy from the surrounding will be initially shock heated to the halo virial temperature. Subsequently, the gas can radiatively cool falling out of hydrostatic equilibrium with the halo. With our estimates of $T \sim 10^5$~K, $n_{\H} \sim 10^{-4}$~{\cc}, and [X/H] $\sim -1.0$, the instantaneous radiative cooling time scale comes out as $t_{cool} \sim 20$~Myrs \citep{Sutherland1993}. These estimates hint at the possibility that the $z = 0.57052$ absorber, with its comparitively low metallicity, could be material infalling into the hot extended corona of the galaxy.

In the case of the $z = 0.61907$ absorber, the SDSS data only suggests that the {\NeVIII} is in a high density region of luminous, massive galaxies. This general picture is consistent with more exhaustive absorber-galaxy surveys that have repeatedly found warm absorbers closely associated with galaxy overdensity regions (e.gs, \citealt{Wakker2009}, \citealt{Chen2009}, \citealt{Narayanan2010b}, \citealt{stocke2014}, \citealt{Werk2016}). Only by extending the galaxy completeness to fainter magnitudes will it be possible to say whether the {\NeVIII} is coincidental with the warm extended envelope of a $L^*$ or fainter galaxy or the shock-heated  gas in intergalactic filaments.

\section{SUMMARY \& CONCLUSIONS}\label{sec9}

We report on the detection of two intervening {\NeVIII} absorbers at $z = 0.61907$ and $z = 0.57052$ in the $HST$/COS far-UV spectrum of the background quasars SDSS~$J080908.13+461925.6$ and SBS~$1122+594$ respectively. The key results are the following. 

\begin{enumerate}

\item The $z = 0.61907$ absorber is seen in {\HI}, {\CII}, {\CIII}, {\NIII}, {\OIII}, {\NIV}, {\OIV}, {\OVI}, and {\NeVIII}. The {\NeVIII}~$770$ line is detected with a significance of $3.0\sigma$ taking into account the statistical and systematic uncertainities. The total column density of the ion is estimated to be $\log~[N(\NeVIII), {\cmsq}] = 13.76~{\pm}~0.14$. The {\NeVIII}~$780$ line is a non-detection. 

\item Ionization models suggest a two-phase solution for the $z = 0.61907$ absorber. The low and intermediate ions are most consistent with photoionized gas with $n_{\H} \sim 7 \times 10^{-4}$~{\cc}, $T \sim 1.5 \times 10^4$~K, $p/K \sim 10.5$~{\cc}~K, and an absorption path length of $L \sim 14.6$~kpc. The ionic column densities yield abundances of [C/H] = [O/H] $= -0.2$, [N/H] $= -1.0$ and [S/H] $= -0.4$, with an uncertainty of $~{\pm}~0.3$~dex. The {\OVI} is also consistent with an origin in this photoionized gas phase. 

\item The {\NeVIII} in the $z = 0.61907$ absorber is found to be tracing a higher temperature gas phase that is dominantly collisionally ionized. The observed $\log~[N(\NeVIII)/N(\OVI) \gtrsim -1.25$ indicates collisional ionization equilibrium temperatures of $T \gtrsim 4.3 \times 10^5$~K for this phase. Using hybrid models of photoionization and collisional ionization, we find this warm phase of the gas to be at $n_{\H} \leq 10^{-3}$~{\cc}, corresponding to an overdensity of $\Delta \sim 1000$.  The {\HI} absorption associated with this gas phase is unknown. Hence the total hydrogen column density, absorber line-of-sight thickness and neon abundance are only approximately constrained to $N(\H) \sim 2.7 \times 10^{20}$~{\cmsq},  $L \sim 87$~kpc and [Ne/H] $\sim -0.5$. 

\item The $z = 0.57052$ absorber has {\HI}, {\NIV}, {\OVI} and {\NeVIII} detected. The {\NeVIII}~$770$ line has a significance of $2.9\sigma$. The $b$-parameters for {\OVI} and the corresponding {\HI} component indicates temperature in the range of $T = (0.5 - 1.5) \times 10^5$~K, which is also consistent with the line widths of {\OIV} and {\NIV}, and supports collisional ionization. 

\item At those warm temperatures, the hybrid models offer a single phase solution with $n_{\H} \sim 10^{-4}$~{\cc} for all the detected ions, corresponding to overdensities of $\Delta \sim 100$. This warm gas phase has a total hydrogen column density of $N(\H) = 4.8 \times 10^{19}$~{\cc}, and a thickness of $L \sim 196$~kpc, with [Ne/H] $\sim$ [O/H] $\sim -1$~dex. 

\item Both warm absorbers are tracing regions where there are a number of galaxies. Over a uniform projected physical separation of $\sim 10$~Mpc and a velocity separation of $\Delta v \sim 1000$~{\kms}, 7 galaxies are identified by SDSS near to the $z = 0.61907$ absorber, and 6 galaxies around the $z = 0.57052$ absorber. In the latter case, the absorber is within the halo virial radius of a $2.6L^*$ galaxy and could be tracing a shock heated gas cloud in the circumgalactic halo of the galaxy. 

\end{enumerate}

Recently, \citet{Hussain2017} remodelled 7 out of the 10 previously known {\NeVIII} absorbers by replacing the earlier UV background models of \citet{HM1996,HM2012} with the KS15. The KS15 radiation field, with the recently updated QSO emissivities and SEDs, has a $\sim 3$ times higher intensity compared to \citet{HM1996,HM2012} at energies $> 4$~Ryd where the ionization potentials of {\OVI} and {\NeVIII} lie. They found that in the absence of any direct evidence for warm-hot temperatures (i.e., the detection of a thermally broad Lyman-$\alpha$ line), the {\NeVIII} can also be explained through photoionization. The KS15 background yields an order of magnitude higher values for density and metallicity compared to the \citet{HM1996}. This results in the absorber line-of-sight thickness decreasing from an unrealistically large $\gtrsim$ Mpc scale to $10 -  200$~kpcs, thereby  making the photoionization solution a viable alternative. In the same models, the photoionized {\NeVIII} gas requires solar, and in most cases supersolar (nearly ten times solar) metallicities to match the observed metal line column densities \citep{Hussain2017}. This is suggestive of the absorption tracing gas that is directly enriched by star formation, and not the canonical low metallicity WHIM distant from galaxies, that simulations predict. On the other hand, the sample of absorbers discussed in this paper have sub-solar metallicities. Interestingly, in the $z = 0.57052$ system for which we have an ionization model independent measure on the warm temperature, the [Ne/H] $\sim - 1.0$~dex, in agreement with the general predictions of \citet{Hussain2017} that at low metallcities, {\NeVIII} can stay longer at high temperature. 

Through cosmological simulations, \citet{Tepper-Garcia2013} had investigated the physical origin of {\NeVIII} systems. Their analysis demonstrates that the peak ionization fraction of $f({\NeVIII}) \sim 10\%$ is achieved through photoionization only at very low densities of $n_{\H} \sim 10^{-6}$~{\cc} (with the KS15 background this changes to $10^{-5}$~{\cc}) and $T \sim 10^4$~K in contrast to collisional ionization ($n_{\H} \sim 10^{-4}$~{\cc}, $T \gtrsim 10^5$~K). As a result, for the same metallicity, the $N(\NeVIII) \propto (n_{\H}~T)^{1/2}$ comes out as $\sim 1 - 2$ orders of magnitude more when it is tracing a warm (collisionaly ionized) phase compared to a photoionized case. Observationally, collisionally ionized {\NeVIII} will be therefore easier to locate in spectra of adequately high sensitivity. Considering this, it is possible that most of the {\NeVIII} detections reported till now (Table \ref{NeVIII_compile}) are likely to be tracers of warm plasma and to a lesser extent photoionized gas. It needs to be mentioned here that the results of \citet{Tepper-Garcia2013} differ from the simulations of \citet{Oppenheimer2012} who find {\NeVIII} to be predominantly from $T \sim 10^4$~K phase of the low-$z$ IGM.  The simulations are inconclusive because of the differences between them in the treatment of physics, the assumptions made and in the implementation of Galactic scale processes. Given such circumstances, the interpretation of observations based on simulations should be done with prudence. 

In certain cases, the data cooperate in a way that allows one to infer the thermal conditions in the absorber without depending on ionization models. For example, in the {\NeVIII} absorbers reported by \citet{savage2005,Savage2011} and \citet{Narayanan2009,Narayanan2012}, the presence of a BLA (albeit at low significance) offered a direct measure on the temperature ($T \sim 5 \times 10^5$~K) of the {\NeVIII} phase of the gas. Similarly, for the z = 0.57052 absorber in this paper, the temperature of the warm plasma was ascertained from the different $b$-values of the metal lines and {\HI}. Identifying BLAs in metal line systems traced by {\OVI} and {\NeVIII} is possibly one of the best ways to infer the presence of warm-hot gas. Detection of thermally broad and shallow {\HI} components in an unambiguous way, particularly in multiphase absorbers, will depend on the availability of high ($S/N \gtrsim 50$) spectroscopic observations with HST /COS.

\section{Acknowledgments}\label{sec10}

SP and AN thank the Department of Space, Government of India, and the Indian Institute of Space Science and Technology for financial support for this work. SM acknowledges support from European Research Council (ERC), Grant Agreement 278594-GasAroundGalaxies. 


\bibliographystyle{mnras}
\bibliography{refer}

\begin{thebibliography}{}
\makeatletter
\relax
\def\mn@urlcharsother{\let\do\@makeother \do\$\do\&\do\#\do\^\do\_\do\%\do\~}
\def\mn@doi{\begingroup\mn@urlcharsother \@ifnextchar [ {\mn@doi@}
  {\mn@doi@[]}}
\def\mn@doi@[#1]#2{\def\@tempa{#1}\ifx\@tempa\@empty \href
  {http://dx.doi.org/#2} {doi:#2}\else \href {http://dx.doi.org/#2} {#1}\fi
  \endgroup}
\def\mn@eprint#1#2{\mn@eprint@#1:#2::\@nil}
\def\mn@eprint@arXiv#1{\href {http://arxiv.org/abs/#1} {{\tt arXiv:#1}}}
\def\mn@eprint@dblp#1{\href {http://dblp.uni-trier.de/rec/bibtex/#1.xml}
  {dblp:#1}}
\def\mn@eprint@#1:#2:#3:#4\@nil{\def\@tempa {#1}\def\@tempb {#2}\def\@tempc
  {#3}\ifx \@tempc \@empty \let \@tempc \@tempb \let \@tempb \@tempa \fi \ifx
  \@tempb \@empty \def\@tempb {arXiv}\fi \@ifundefined
  {mn@eprint@\@tempb}{\@tempb:\@tempc}{\expandafter \expandafter \csname
  mn@eprint@\@tempb\endcsname \expandafter{\@tempc}}}

\bibitem[\protect\citeauthoryear{{Asplund}, {Grevesse}, {Sauval}  \&
  {Scott}}{{Asplund} et~al.}{2009}]{Asplund2009}
{Asplund} M.,  {Grevesse} N.,  {Sauval} A.~J.,   {Scott} P.,  2009, \mn@doi
  [\araa] {10.1146/annurev.astro.46.060407.145222}, \href
  {http://adsabs.harvard.edu/abs/2009ARA26A..47..481A} {47, 481}

\bibitem[\protect\citeauthoryear{{Birnboim} \& {Dekel}}{{Birnboim} \&
  {Dekel}}{2003}]{Birnboim2003}
{Birnboim} Y.,  {Dekel} A.,  2003, \mn@doi [\mnras]
  {10.1046/j.1365-8711.2003.06955.x}, \href
  {http://adsabs.harvard.edu/abs/2003MNRAS.345..349B} {345, 349}

\bibitem[\protect\citeauthoryear{{Bolton} \& {Becker}}{{Bolton} \&
  {Becker}}{2009}]{Bolton2009}
{Bolton} J.~S.,  {Becker} G.~D.,  2009, \mn@doi [\mnras]
  {10.1111/j.1745-3933.2009.00700.x}, \href
  {http://adsabs.harvard.edu/abs/2009MNRAS.398L..26B} {398, L26}

\bibitem[\protect\citeauthoryear{{Bordoloi} et~al.,}{{Bordoloi}
  et~al.}{2014}]{Bordoloi2014}
{Bordoloi} R.,  et~al., 2014, \mn@doi [\apj] {10.1088/0004-637X/796/2/136},
  \href {http://adsabs.harvard.edu/abs/2014ApJ...796..136B} {796, 136}

\bibitem[\protect\citeauthoryear{{Bordoloi}, {Heckman}  \& {Norman}}{{Bordoloi}
  et~al.}{2016}]{Bordoloi2016}
{Bordoloi} R.,  {Heckman} T.~M.,   {Norman} C.~A.,  2016, preprint, \href
  {http://adsabs.harvard.edu/abs/2016arXiv160507187B} {} (\mn@eprint {arXiv}
  {1605.07187})

\bibitem[\protect\citeauthoryear{{Cen} \& {Ostriker}}{{Cen} \&
  {Ostriker}}{1999}]{Cen1999}
{Cen} R.,  {Ostriker} J.~P.,  1999, \mn@doi [\apj] {10.1086/306949}, \href
  {http://adsabs.harvard.edu/abs/1999ApJ...514....1C} {514, 1}

\bibitem[\protect\citeauthoryear{{Cen} \& {Ostriker}}{{Cen} \&
  {Ostriker}}{2006}]{Cen2006}
{Cen} R.,  {Ostriker} J.~P.,  2006, \mn@doi [\apj] {10.1086/506505}, \href
  {http://adsabs.harvard.edu/abs/2006ApJ...650..560C} {650, 560}

\bibitem[\protect\citeauthoryear{{Cen}, {Miralda-Escud{\'e}}, {Ostriker}  \&
  {Rauch}}{{Cen} et~al.}{1994}]{Cen1994}
{Cen} R.,  {Miralda-Escud{\'e}} J.,  {Ostriker} J.~P.,   {Rauch} M.,  1994,
  \mn@doi [\apjl] {10.1086/187670}, \href
  {http://adsabs.harvard.edu/abs/1994ApJ...437L...9C} {437, L9}

\bibitem[\protect\citeauthoryear{{Chen} \& {Mulchaey}}{{Chen} \&
  {Mulchaey}}{2009}]{Chen2009}
{Chen} H.-W.,  {Mulchaey} J.~S.,  2009, \mn@doi [\apj]
  {10.1088/0004-637X/701/2/1219}, \href
  {http://adsabs.harvard.edu/abs/2009ApJ...701.1219C} {701, 1219}

\bibitem[\protect\citeauthoryear{{Croom} et~al.,}{{Croom}
  et~al.}{2009}]{Croom2009}
{Croom} S.~M.,  et~al., 2009, \mn@doi [\mnras]
  {10.1111/j.1365-2966.2009.15398.x}, \href
  {http://adsabs.harvard.edu/abs/2009MNRAS.399.1755C} {399, 1755}

\bibitem[\protect\citeauthoryear{{Danforth} \& {Shull}}{{Danforth} \&
  {Shull}}{2005}]{Danforth2005}
{Danforth} C.~W.,  {Shull} J.~M.,  2005, \mn@doi [\apj] {10.1086/429285}, \href
  {http://adsabs.harvard.edu/abs/2005ApJ...624..555D} {624, 555}

\bibitem[\protect\citeauthoryear{{Danforth} \& {Shull}}{{Danforth} \&
  {Shull}}{2008}]{Danforth2008}
{Danforth} C.~W.,  {Shull} J.~M.,  2008, \mn@doi [\apj] {10.1086/587127}, \href
  {http://adsabs.harvard.edu/abs/2008ApJ...679..194D} {679, 194}

\bibitem[\protect\citeauthoryear{{Danforth}, {Keeney}, {Stocke}, {Shull}  \&
  {Yao}}{{Danforth} et~al.}{2010}]{Danforth2010}
{Danforth} C.~W.,  {Keeney} B.~A.,  {Stocke} J.~T.,  {Shull} J.~M.,   {Yao} Y.,
   2010, \mn@doi [\apj] {10.1088/0004-637X/720/1/976}, \href
  {http://adsabs.harvard.edu/abs/2010ApJ...720..976D} {720, 976}

\bibitem[\protect\citeauthoryear{{Danforth} et~al.,}{{Danforth}
  et~al.}{2016}]{Danforth2016}
{Danforth} C.~W.,  et~al., 2016, \mn@doi [\apj] {10.3847/0004-637X/817/2/111},
  \href {http://adsabs.harvard.edu/abs/2016ApJ...817..111D} {817, 111}

\bibitem[\protect\citeauthoryear{{Dav{\'e}} et~al.,}{{Dav{\'e}}
  et~al.}{2001}]{Dave2001}
{Dav{\'e}} R.,  et~al., 2001, \mn@doi [\apj] {10.1086/320548}, \href
  {http://adsabs.harvard.edu/abs/2001ApJ...552..473D} {552, 473}

\bibitem[\protect\citeauthoryear{{Dixon}}{{Dixon}}{2010}]{Dixon2010}
{Dixon} W.,  2010, {STIS CCD Full-Field Sensitivity Monitor C18}, HST Proposal

\bibitem[\protect\citeauthoryear{{Drake} \& {Testa}}{{Drake} \&
  {Testa}}{2005}]{Drake2005}
{Drake} J.~J.,  {Testa} P.,  2005, \mn@doi [\nat] {10.1038/nature03803}, \href
  {http://adsabs.harvard.edu/abs/2005Natur.436..525D} {436, 525}

\bibitem[\protect\citeauthoryear{{Ferland} et~al.,}{{Ferland}
  et~al.}{2013}]{2013RMxAA..49..137F}
{Ferland} G.~J.,  et~al., 2013, \rmxaa, \href
  {http://adsabs.harvard.edu/abs/2013RMxAA..49..137F} {49, 137}

\bibitem[\protect\citeauthoryear{{Foltz}, {Weymann}, {Peterson}, {Sun},
  {Malkan}  \& {Chaffee}}{{Foltz} et~al.}{1986}]{Foltz1986}
{Foltz} C.~B.,  {Weymann} R.~J.,  {Peterson} B.~M.,  {Sun} L.,  {Malkan} M.~A.,
    {Chaffee} Jr. F.~H.,  1986, \mn@doi [\apj] {10.1086/164440}, \href
  {http://adsabs.harvard.edu/abs/1986ApJ...307..504F} {307, 504}

\bibitem[\protect\citeauthoryear{{Fox}, {Ledoux}, {Vreeswijk}, {Smette}  \&
  {Jaunsen}}{{Fox} et~al.}{2008}]{Fox2008}
{Fox} A.~J.,  {Ledoux} C.,  {Vreeswijk} P.~M.,  {Smette} A.,   {Jaunsen} A.~O.,
   2008, \mn@doi [\aap] {10.1051/0004-6361:200810286}, \href
  {http://adsabs.harvard.edu/abs/2008A%26A...491..189F} {491, 189}

\bibitem[\protect\citeauthoryear{{Froning} \& {Green}}{{Froning} \&
  {Green}}{2009}]{Froning2009}
{Froning} C.~S.,  {Green} J.~C.,  2009, \mn@doi [\apss]
  {10.1007/s10509-008-9758-y}, \href
  {http://adsabs.harvard.edu/abs/2009Ap%26SS.320..181F} {320, 181}

\bibitem[\protect\citeauthoryear{{Fukugita} \& {Peebles}}{{Fukugita} \&
  {Peebles}}{2006}]{Fukugita2006}
{Fukugita} M.,  {Peebles} P.~J.~E.,  2006, \mn@doi [\apj] {10.1086/499556},
  \href {http://adsabs.harvard.edu/abs/2006ApJ...639..590F} {639, 590}

\bibitem[\protect\citeauthoryear{{Gaikwad}, {Khaire}, {Choudhury}  \&
  {Srianand}}{{Gaikwad} et~al.}{2017a}]{Gaikwad2017a}
{Gaikwad} P.,  {Khaire} V.,  {Choudhury} T.~R.,   {Srianand} R.,  2017a,
  \mn@doi [\mnras] {10.1093/mnras/stw3086}, \href
  {http://adsabs.harvard.edu/abs/2017MNRAS.466..838G} {466, 838}

\bibitem[\protect\citeauthoryear{{Gaikwad}, {Srianand}, {Choudhury}  \&
  {Khaire}}{{Gaikwad} et~al.}{2017b}]{Gaikwad2017b}
{Gaikwad} P.,  {Srianand} R.,  {Choudhury} T.~R.,   {Khaire} V.,  2017b,
  \mn@doi [\mnras] {10.1093/mnras/stx248}, \href
  {http://adsabs.harvard.edu/abs/2017MNRAS.467.3172G} {467, 3172}

\bibitem[\protect\citeauthoryear{{Ganguly}, {Sembach}, {Tripp}, {Savage}  \&
  {Wakker}}{{Ganguly} et~al.}{2006}]{Ganguly2006}
{Ganguly} R.,  {Sembach} K.~R.,  {Tripp} T.~M.,  {Savage} B.~D.,   {Wakker}
  B.~P.,  2006, \mn@doi [\apj] {10.1086/504395}, \href
  {http://adsabs.harvard.edu/abs/2006ApJ...645..868G} {645, 868}

\bibitem[\protect\citeauthoryear{{Gnat} \& {Sternberg}}{{Gnat} \&
  {Sternberg}}{2007}]{Gnat2007}
{Gnat} O.,  {Sternberg} A.,  2007, \mn@doi [\apjs] {10.1086/509786}, \href
  {http://adsabs.harvard.edu/abs/2007ApJS..168..213G} {168, 213}

\bibitem[\protect\citeauthoryear{{Green} et~al.,}{{Green}
  et~al.}{2012}]{Green2012}
{Green} J.~C.,  et~al., 2012, \mn@doi [\apj] {10.1088/0004-637X/744/1/60},
  \href {http://adsabs.harvard.edu/abs/2012ApJ...744...60G} {744, 60}

\bibitem[\protect\citeauthoryear{{Haardt} \& {Madau}}{{Haardt} \&
  {Madau}}{1996}]{HM1996}
{Haardt} F.,  {Madau} P.,  1996, \mn@doi [\apj] {10.1086/177035}, \href
  {http://adsabs.harvard.edu/abs/1996ApJ...461...20H} {461, 20}

\bibitem[\protect\citeauthoryear{{Haardt} \& {Madau}}{{Haardt} \&
  {Madau}}{2012}]{HM2012}
{Haardt} F.,  {Madau} P.,  2012, \mn@doi [\apj] {10.1088/0004-637X/746/2/125},
  \href {http://adsabs.harvard.edu/abs/2012ApJ...746..125H} {746, 125}

\bibitem[\protect\citeauthoryear{{Hernquist}, {Katz}, {Weinberg}  \&
  {Miralda-Escud{\'e}}}{{Hernquist} et~al.}{1996}]{Hernquist1996}
{Hernquist} L.,  {Katz} N.,  {Weinberg} D.~H.,   {Miralda-Escud{\'e}} J.,
  1996, \mn@doi [\apjl] {10.1086/309899}, \href
  {http://adsabs.harvard.edu/abs/1996ApJ...457L..51H} {457, L51}

\bibitem[\protect\citeauthoryear{{Hewett} \& {Wild}}{{Hewett} \&
  {Wild}}{2010}]{Hewett2010}
{Hewett} P.~C.,  {Wild} V.,  2010, \mn@doi [\mnras]
  {10.1111/j.1365-2966.2010.16648.x}, \href
  {http://adsabs.harvard.edu/abs/2010MNRAS.405.2302H} {405, 2302}

\bibitem[\protect\citeauthoryear{{Howk}, {Savage}, {Sembach}  \&
  {Hoopes}}{{Howk} et~al.}{2002}]{Howk2002}
{Howk} J.~C.,  {Savage} B.~D.,  {Sembach} K.~R.,   {Hoopes} C.~G.,  2002,
  \mn@doi [\apj] {10.1086/340231}, \href
  {http://adsabs.harvard.edu/abs/2002ApJ...572..264H} {572, 264}

\bibitem[\protect\citeauthoryear{{Hussain}, {Muzahid}, {Narayanan}, {Srianand},
  {Wakker}, {Charlton}  \& {Pathak}}{{Hussain} et~al.}{2015}]{Hussain2015}
{Hussain} T.,  {Muzahid} S.,  {Narayanan} A.,  {Srianand} R.,  {Wakker} B.~P.,
  {Charlton} J.~C.,   {Pathak} A.,  2015, \mn@doi [\mnras]
  {10.1093/mnras/stu2285}, \href
  {http://adsabs.harvard.edu/abs/2015MNRAS.446.2444H} {446, 2444}

\bibitem[\protect\citeauthoryear{{Hussain}, {Khaire}, {Srianand}, {Muzahid}  \&
  {Pathak}}{{Hussain} et~al.}{2017}]{Hussain2017}
{Hussain} T.,  {Khaire} V.,  {Srianand} R.,  {Muzahid} S.,   {Pathak} A.,
  2017, \mn@doi [\mnras] {10.1093/mnras/stw3265}, \href
  {http://adsabs.harvard.edu/abs/2017MNRAS.466.3133H} {466, 3133}

\bibitem[\protect\citeauthoryear{{Ilbert} et~al.,}{{Ilbert}
  et~al.}{2005}]{Ilbert2005}
{Ilbert} O.,  et~al., 2005, \mn@doi [\aap] {10.1051/0004-6361:20041961}, \href
  {http://adsabs.harvard.edu/abs/2005A%26A...439..863I} {439, 863}

\bibitem[\protect\citeauthoryear{{Khaire} \& {Srianand}}{{Khaire} \&
  {Srianand}}{2015a}]{KS2015b}
{Khaire} V.,  {Srianand} R.,  2015a, \mn@doi [\mnras] {10.1093/mnrasl/slv060},
  \href {http://adsabs.harvard.edu/abs/2015MNRAS.451L..30K} {451, L30}

\bibitem[\protect\citeauthoryear{{Khaire} \& {Srianand}}{{Khaire} \&
  {Srianand}}{2015b}]{KS2015a}
{Khaire} V.,  {Srianand} R.,  2015b, \mn@doi [\apj]
  {10.1088/0004-637X/805/1/33}, \href
  {http://adsabs.harvard.edu/abs/2015ApJ...805...33K} {805, 33}

\bibitem[\protect\citeauthoryear{{Kollmeier} et~al.,}{{Kollmeier}
  et~al.}{2014}]{Kollmeier2014}
{Kollmeier} J.~A.,  et~al., 2014, \mn@doi [\apjl]
  {10.1088/2041-8205/789/2/L32}, \href
  {http://adsabs.harvard.edu/abs/2014ApJ...789L..32K} {789, L32}

\bibitem[\protect\citeauthoryear{{Kriss}}{{Kriss}}{2011}]{Kriss2011}
{Kriss} G.~A.,  2011, Technical report, {Improved Medium Resolution Line Spread
  Functions for COS FUV Spectra}

\bibitem[\protect\citeauthoryear{{Lehner}, {Prochaska}, {Kobulnicky},
  {Cooksey}, {Howk}, {Williger}  \& {Cales}}{{Lehner}
  et~al.}{2009}]{Lehner2009}
{Lehner} N.,  {Prochaska} J.~X.,  {Kobulnicky} H.~A.,  {Cooksey} K.~L.,  {Howk}
  J.~C.,  {Williger} G.~M.,   {Cales} S.~L.,  2009, \mn@doi [\apj]
  {10.1088/0004-637X/694/2/734}, \href
  {http://adsabs.harvard.edu/abs/2009ApJ...694..734L} {694, 734}

\bibitem[\protect\citeauthoryear{{Maller} \& {Bullock}}{{Maller} \&
  {Bullock}}{2004}]{Maller2004}
{Maller} A.~H.,  {Bullock} J.~S.,  2004, \mn@doi [\mnras]
  {10.1111/j.1365-2966.2004.08349.x}, \href
  {http://adsabs.harvard.edu/abs/2004MNRAS.355..694M} {355, 694}

\bibitem[\protect\citeauthoryear{{Meiring}, {Tripp}, {Werk}, {Howk}, {Jenkins},
  {Prochaska}, {Lehner}  \& {Sembach}}{{Meiring} et~al.}{2013}]{Meiring2013}
{Meiring} J.~D.,  {Tripp} T.~M.,  {Werk} J.~K.,  {Howk} J.~C.,  {Jenkins}
  E.~B.,  {Prochaska} J.~X.,  {Lehner} N.,   {Sembach} K.~R.,  2013, \mn@doi
  [\apj] {10.1088/0004-637X/767/1/49}, \href
  {http://adsabs.harvard.edu/abs/2013ApJ...767...49M} {767, 49}

\bibitem[\protect\citeauthoryear{{Miralda-Escude} \&
  {Ostriker}}{{Miralda-Escude} \& {Ostriker}}{1990}]{Escude1990}
{Miralda-Escude} J.,  {Ostriker} J.~P.,  1990, \mn@doi [\apj] {10.1086/168358},
  \href {http://adsabs.harvard.edu/abs/1990ApJ...350....1M} {350, 1}

\bibitem[\protect\citeauthoryear{{Miralda-Escud{\'e}}, {Cen}, {Ostriker}  \&
  {Rauch}}{{Miralda-Escud{\'e}} et~al.}{1996}]{Escude1996}
{Miralda-Escud{\'e}} J.,  {Cen} R.,  {Ostriker} J.~P.,   {Rauch} M.,  1996,
  \mn@doi [\apj] {10.1086/177992}, \href
  {http://adsabs.harvard.edu/abs/1996ApJ...471..582M} {471, 582}

\bibitem[\protect\citeauthoryear{{Morton}}{{Morton}}{2003}]{morton2003}
{Morton} D.~C.,  2003, \mn@doi [\apjs] {10.1086/377639}, \href
  {http://adsabs.harvard.edu/abs/2003ApJS..149..205M} {149, 205}

\bibitem[\protect\citeauthoryear{{Mulchaey} \& {Chen}}{{Mulchaey} \&
  {Chen}}{2009}]{Mulchaey2009}
{Mulchaey} J.~S.,  {Chen} H.-W.,  2009, \mn@doi [\apjl]
  {10.1088/0004-637X/698/1/L46}, \href
  {http://adsabs.harvard.edu/abs/2009ApJ...698L..46M} {698, L46}

\bibitem[\protect\citeauthoryear{{Muzahid}}{{Muzahid}}{2014}]{Muzahid2014}
{Muzahid} S.,  2014, \mn@doi [\apj] {10.1088/0004-637X/784/1/5}, \href
  {http://adsabs.harvard.edu/abs/2014ApJ...784....5M} {784, 5}

\bibitem[\protect\citeauthoryear{{Muzahid}, {Srianand}, {Savage}, {Narayanan},
  {Mohan}  \& {Dewangan}}{{Muzahid} et~al.}{2012}]{Muzahid2012}
{Muzahid} S.,  {Srianand} R.,  {Savage} B.~D.,  {Narayanan} A.,  {Mohan} V.,
  {Dewangan} G.~C.,  2012, \mn@doi [\mnras] {10.1111/j.1745-3933.2012.01288.x},
  \href {http://adsabs.harvard.edu/abs/2012MNRAS.424L..59M} {424, L59}

\bibitem[\protect\citeauthoryear{{Muzahid}, {Srianand}, {Arav}, {Savage}  \&
  {Narayanan}}{{Muzahid} et~al.}{2013}]{Muzahid2013}
{Muzahid} S.,  {Srianand} R.,  {Arav} N.,  {Savage} B.~D.,   {Narayanan} A.,
  2013, \mn@doi [\mnras] {10.1093/mnras/stt390}, \href
  {http://adsabs.harvard.edu/abs/2013MNRAS.431.2885M} {431, 2885}

\bibitem[\protect\citeauthoryear{{Narayanan}, {Wakker}  \&
  {Savage}}{{Narayanan} et~al.}{2009}]{Narayanan2009}
{Narayanan} A.,  {Wakker} B.~P.,   {Savage} B.~D.,  2009, \mn@doi [\apj]
  {10.1088/0004-637X/703/1/74}, \href
  {http://adsabs.harvard.edu/abs/2009ApJ...703...74N} {703, 74}

\bibitem[\protect\citeauthoryear{{Narayanan}, {Savage}  \&
  {Wakker}}{{Narayanan} et~al.}{2010a}]{Narayanan2010a}
{Narayanan} A.,  {Savage} B.~D.,   {Wakker} B.~P.,  2010a, \mn@doi [\apj]
  {10.1088/0004-637X/712/2/1443}, \href
  {http://adsabs.harvard.edu/abs/2010ApJ...712.1443N} {712, 1443}

\bibitem[\protect\citeauthoryear{{Narayanan}, {Wakker}, {Savage}, {Keeney},
  {Shull}, {Stocke}  \& {Sembach}}{{Narayanan} et~al.}{2010b}]{Narayanan2010b}
{Narayanan} A.,  {Wakker} B.~P.,  {Savage} B.~D.,  {Keeney} B.~A.,  {Shull}
  J.~M.,  {Stocke} J.~T.,   {Sembach} K.~R.,  2010b, \mn@doi [\apj]
  {10.1088/0004-637X/721/2/960}, \href
  {http://adsabs.harvard.edu/abs/2010ApJ...721..960N} {721, 960}

\bibitem[\protect\citeauthoryear{{Narayanan} et~al.,}{{Narayanan}
  et~al.}{2011}]{Narayanan2011a}
{Narayanan} A.,  et~al., 2011, \mn@doi [\apj] {10.1088/0004-637X/730/1/15},
  \href {http://adsabs.harvard.edu/abs/2011ApJ...730...15N} {730, 15}

\bibitem[\protect\citeauthoryear{{Narayanan}, {Savage}  \&
  {Wakker}}{{Narayanan} et~al.}{2012}]{Narayanan2012}
{Narayanan} A.,  {Savage} B.~D.,   {Wakker} B.~P.,  2012, \mn@doi [\apj]
  {10.1088/0004-637X/752/1/65}, \href
  {http://adsabs.harvard.edu/abs/2012ApJ...752...65N} {752, 65}

\bibitem[\protect\citeauthoryear{{Oppenheimer} \& {Dav{\'e}}}{{Oppenheimer} \&
  {Dav{\'e}}}{2009}]{Oppenheimer2009}
{Oppenheimer} B.~D.,  {Dav{\'e}} R.,  2009, \mn@doi [\mnras]
  {10.1111/j.1365-2966.2009.14676.x}, \href
  {http://adsabs.harvard.edu/abs/2009MNRAS.395.1875O} {395, 1875}

\bibitem[\protect\citeauthoryear{{Oppenheimer}, {Dav{\'e}}, {Katz}, {Kollmeier}
   \& {Weinberg}}{{Oppenheimer} et~al.}{2012}]{Oppenheimer2012}
{Oppenheimer} B.~D.,  {Dav{\'e}} R.,  {Katz} N.,  {Kollmeier} J.~A.,
  {Weinberg} D.~H.,  2012, \mn@doi [\mnras] {10.1111/j.1365-2966.2011.20096.x},
  \href {http://adsabs.harvard.edu/abs/2012MNRAS.420..829O} {420, 829}

\bibitem[\protect\citeauthoryear{{Oppenheimer} et~al.,}{{Oppenheimer}
  et~al.}{2016}]{Oppenheimer2016}
{Oppenheimer} B.~D.,  et~al., 2016, \mn@doi [\mnras] {10.1093/mnras/stw1066},
  \href {http://adsabs.harvard.edu/abs/2016MNRAS.460.2157O} {460, 2157}

\bibitem[\protect\citeauthoryear{{Osterman} et~al.,}{{Osterman}
  et~al.}{2011}]{Osterman2011}
{Osterman} S.,  et~al., 2011, \mn@doi [\apss] {10.1007/s10509-011-0699-5},
  \href {http://adsabs.harvard.edu/abs/2011Ap%26SS.335..257O} {335, 257}

\bibitem[\protect\citeauthoryear{{Pachat}, {Narayanan}, {Muzahid}, {Khaire},
  {Srianand}, {Wakker}  \& {Savage}}{{Pachat} et~al.}{2016}]{Pachat2016}
{Pachat} S.,  {Narayanan} A.,  {Muzahid} S.,  {Khaire} V.,  {Srianand} R.,
  {Wakker} B.~P.,   {Savage} B.~D.,  2016, \mn@doi [\mnras]
  {10.1093/mnras/stw194}, \href
  {http://adsabs.harvard.edu/abs/2016MNRAS.458..733P} {458, 733}

\bibitem[\protect\citeauthoryear{{Palanque-Delabrouille}
  et~al.,}{{Palanque-Delabrouille} et~al.}{2013}]{PD2013}
{Palanque-Delabrouille} N.,  et~al., 2013, \mn@doi [\aap]
  {10.1051/0004-6361/201220379}, \href
  {http://adsabs.harvard.edu/abs/2013A%26A...551A..29P} {551, A29}

\bibitem[\protect\citeauthoryear{{Persic} \& {Salucci}}{{Persic} \&
  {Salucci}}{1992}]{Persic1992}
{Persic} M.,  {Salucci} P.,  1992, \mn@doi [\mnras] {10.1093/mnras/258.1.14P},
  \href {http://adsabs.harvard.edu/abs/1992MNRAS.258P..14P} {258, 14P}

\bibitem[\protect\citeauthoryear{{Petitjean} \& {Srianand}}{{Petitjean} \&
  {Srianand}}{1999}]{Petitjean1999}
{Petitjean} P.,  {Srianand} R.,  1999, \aap, \href
  {http://adsabs.harvard.edu/abs/1999A%26A...345...73P} {345, 73}

\bibitem[\protect\citeauthoryear{{Qu} \& {Bregman}}{{Qu} \&
  {Bregman}}{2016}]{Qu2016}
{Qu} Z.,  {Bregman} J.~N.,  2016, \mn@doi [\apj] {10.3847/0004-637X/832/2/189},
  \href {http://adsabs.harvard.edu/abs/2016ApJ...832..189Q} {832, 189}

\bibitem[\protect\citeauthoryear{{Rauch}}{{Rauch}}{1998}]{Rauch1998}
{Rauch} M.,  1998, \mn@doi [\araa] {10.1146/annurev.astro.36.1.267}, \href
  {http://adsabs.harvard.edu/abs/1998ARA26A..36..267R} {36, 267}

\bibitem[\protect\citeauthoryear{{Rauch} et~al.,}{{Rauch}
  et~al.}{1997}]{Rauch1997}
{Rauch} M.,  et~al., 1997, \mn@doi [\apj] {10.1086/304765}, \href
  {http://adsabs.harvard.edu/abs/1997ApJ...489....7R} {489, 7}

\bibitem[\protect\citeauthoryear{{Savage} \& {Sembach}}{{Savage} \&
  {Sembach}}{1991}]{savage1991}
{Savage} B.~D.,  {Sembach} K.~R.,  1991, \mn@doi [\apj] {10.1086/170498}, \href
  {http://adsabs.harvard.edu/abs/1991ApJ...379..245S} {379, 245}

\bibitem[\protect\citeauthoryear{{Savage}, {Sembach}, {Tripp}  \&
  {Richter}}{{Savage} et~al.}{2002}]{Savage2002}
{Savage} B.~D.,  {Sembach} K.~R.,  {Tripp} T.~M.,   {Richter} P.,  2002,
  \mn@doi [\apj] {10.1086/324288}, \href
  {http://adsabs.harvard.edu/abs/2002ApJ...564..631S} {564, 631}

\bibitem[\protect\citeauthoryear{{Savage}, {Lehner}, {Wakker}, {Sembach}  \&
  {Tripp}}{{Savage} et~al.}{2005}]{savage2005}
{Savage} B.~D.,  {Lehner} N.,  {Wakker} B.~P.,  {Sembach} K.~R.,   {Tripp}
  T.~M.,  2005, \mn@doi [\apj] {10.1086/429985}, \href
  {http://adsabs.harvard.edu/abs/2005ApJ...626..776S} {626, 776}

\bibitem[\protect\citeauthoryear{{Savage}, {Lehner}  \& {Narayanan}}{{Savage}
  et~al.}{2011}]{Savage2011}
{Savage} B.~D.,  {Lehner} N.,   {Narayanan} A.,  2011, \mn@doi [\apj]
  {10.1088/0004-637X/743/2/180}, \href
  {http://adsabs.harvard.edu/abs/2011ApJ...743..180S} {743, 180}

\bibitem[\protect\citeauthoryear{{Savage}, {Kim}, {Wakker}, {Keeney}, {Shull},
  {Stocke}  \& {Green}}{{Savage} et~al.}{2014}]{Savage2014}
{Savage} B.~D.,  {Kim} T.-S.,  {Wakker} B.~P.,  {Keeney} B.,  {Shull} J.~M.,
  {Stocke} J.~T.,   {Green} J.~C.,  2014, \mn@doi [\apjs]
  {10.1088/0067-0049/212/1/8}, \href
  {http://adsabs.harvard.edu/abs/2014ApJS..212....8S} {212, 8}

\bibitem[\protect\citeauthoryear{{Shen} \& {M{\'e}nard}}{{Shen} \&
  {M{\'e}nard}}{2012}]{Shen2012}
{Shen} Y.,  {M{\'e}nard} B.,  2012, \mn@doi [\apj]
  {10.1088/0004-637X/748/2/131}, \href
  {http://adsabs.harvard.edu/abs/2012ApJ...748..131S} {748, 131}

\bibitem[\protect\citeauthoryear{{Shull}, {Roberts}, {Giroux}, {Penton}  \&
  {Fardal}}{{Shull} et~al.}{1999}]{Shull1999}
{Shull} J.~M.,  {Roberts} D.,  {Giroux} M.~L.,  {Penton} S.~V.,   {Fardal}
  M.~A.,  1999, \mn@doi [\aj] {10.1086/301053}, \href
  {http://adsabs.harvard.edu/abs/1999AJ....118.1450S} {118, 1450}

\bibitem[\protect\citeauthoryear{{Shull}, {Moloney}, {Danforth}  \&
  {Tilton}}{{Shull} et~al.}{2015}]{Shull2015}
{Shull} J.~M.,  {Moloney} J.,  {Danforth} C.~W.,   {Tilton} E.~M.,  2015,
  \mn@doi [\apj] {10.1088/0004-637X/811/1/3}, \href
  {http://adsabs.harvard.edu/abs/2015ApJ...811....3S} {811, 3}

\bibitem[\protect\citeauthoryear{{Smith}, {Hallman}, {Shull}  \&
  {O'Shea}}{{Smith} et~al.}{2011}]{Smith2011}
{Smith} B.~D.,  {Hallman} E.~J.,  {Shull} J.~M.,   {O'Shea} B.~W.,  2011,
  \mn@doi [\apj] {10.1088/0004-637X/731/1/6}, \href
  {http://adsabs.harvard.edu/abs/2011ApJ...731....6S} {731, 6}

\bibitem[\protect\citeauthoryear{{Stevans}, {Shull}, {Danforth}  \&
  {Tilton}}{{Stevans} et~al.}{2014}]{stevans2014}
{Stevans} M.~L.,  {Shull} J.~M.,  {Danforth} C.~W.,   {Tilton} E.~M.,  2014,
  \mn@doi [\apj] {10.1088/0004-637X/794/1/75}, \href
  {http://adsabs.harvard.edu/abs/2014ApJ...794...75S} {794, 75}

\bibitem[\protect\citeauthoryear{{Stocke} et~al.,}{{Stocke}
  et~al.}{2014}]{stocke2014}
{Stocke} J.~T.,  et~al., 2014, \mn@doi [\apj] {10.1088/0004-637X/791/2/128},
  \href {http://adsabs.harvard.edu/abs/2014ApJ...791..128S} {791, 128}

\bibitem[\protect\citeauthoryear{{Strauss} et~al.,}{{Strauss}
  et~al.}{2002}]{Strauss2002}
{Strauss} M.~A.,  et~al., 2002, \mn@doi [\aj] {10.1086/342343}, \href
  {http://adsabs.harvard.edu/abs/2002AJ....124.1810S} {124, 1810}

\bibitem[\protect\citeauthoryear{{Sutherland} \& {Dopita}}{{Sutherland} \&
  {Dopita}}{1993}]{Sutherland1993}
{Sutherland} R.~S.,  {Dopita} M.~A.,  1993, \mn@doi [\apjs] {10.1086/191823},
  \href {http://adsabs.harvard.edu/abs/1993ApJS...88..253S} {88, 253}

\bibitem[\protect\citeauthoryear{{Tepper-Garc{\'{\i}}a}, {Richter}, {Schaye},
  {Booth}, {Dalla Vecchia}, {Theuns}  \& {Wiersma}}{{Tepper-Garc{\'{\i}}a}
  et~al.}{2011}]{Tepper-Garcia2011}
{Tepper-Garc{\'{\i}}a} T.,  {Richter} P.,  {Schaye} J.,  {Booth} C.~M.,  {Dalla
  Vecchia} C.,  {Theuns} T.,   {Wiersma} R.~P.~C.,  2011, \mn@doi [\mnras]
  {10.1111/j.1365-2966.2010.18123.x}, \href
  {http://adsabs.harvard.edu/abs/2011MNRAS.413..190T} {413, 190}

\bibitem[\protect\citeauthoryear{{Tepper-Garc{\'{\i}}a}, {Richter}  \&
  {Schaye}}{{Tepper-Garc{\'{\i}}a} et~al.}{2013}]{Tepper-Garcia2013}
{Tepper-Garc{\'{\i}}a} T.,  {Richter} P.,   {Schaye} J.,  2013, \mn@doi
  [\mnras] {10.1093/mnras/stt1712}, \href
  {http://adsabs.harvard.edu/abs/2013MNRAS.436.2063T} {436, 2063}

\bibitem[\protect\citeauthoryear{{Thom} \& {Chen}}{{Thom} \&
  {Chen}}{2008a}]{Thom2008}
{Thom} C.,  {Chen} H.-W.,  2008a, \mn@doi [\apjs] {10.1086/591232}, \href
  {http://adsabs.harvard.edu/abs/2008ApJS..179...37T} {179, 37}

\bibitem[\protect\citeauthoryear{{Thom} \& {Chen}}{{Thom} \&
  {Chen}}{2008b}]{Thom2008b}
{Thom} C.,  {Chen} H.-W.,  2008b, \mn@doi [\apj] {10.1086/587976}, \href
  {http://adsabs.harvard.edu/abs/2008ApJ...683...22T} {683, 22}

\bibitem[\protect\citeauthoryear{{Tinker} \& {Conroy}}{{Tinker} \&
  {Conroy}}{2009}]{Tinker2009}
{Tinker} J.~L.,  {Conroy} C.,  2009, \mn@doi [\apj]
  {10.1088/0004-637X/691/1/633}, \href
  {http://adsabs.harvard.edu/abs/2009ApJ...691..633T} {691, 633}

\bibitem[\protect\citeauthoryear{{Tripp}, {Savage}  \& {Jenkins}}{{Tripp}
  et~al.}{2000}]{Tripp2000}
{Tripp} T.~M.,  {Savage} B.~D.,   {Jenkins} E.~B.,  2000, \mn@doi [\apjl]
  {10.1086/312644}, \href {http://adsabs.harvard.edu/abs/2000ApJ...534L...1T}
  {534, L1}

\bibitem[\protect\citeauthoryear{{Tripp}, {Sembach}, {Bowen}, {Savage},
  {Jenkins}, {Lehner}  \& {Richter}}{{Tripp} et~al.}{2008}]{Tripp2008}
{Tripp} T.~M.,  {Sembach} K.~R.,  {Bowen} D.~V.,  {Savage} B.~D.,  {Jenkins}
  E.~B.,  {Lehner} N.,   {Richter} P.,  2008, \mn@doi [\apjs] {10.1086/587486},
  \href {http://adsabs.harvard.edu/abs/2008ApJS..177...39T} {177, 39}

\bibitem[\protect\citeauthoryear{{Tripp} et~al.,}{{Tripp}
  et~al.}{2011}]{Tripp2011}
{Tripp} T.~M.,  et~al., 2011, \mn@doi [Science] {10.1126/science.1209850},
  \href {http://adsabs.harvard.edu/abs/2011Sci...334..952T} {334, 952}

\bibitem[\protect\citeauthoryear{{Tumlinson} et~al.,}{{Tumlinson}
  et~al.}{2011}]{Tumlinson2011}
{Tumlinson} J.,  et~al., 2011, \mn@doi [\apj] {10.1088/0004-637X/733/2/111},
  \href {http://adsabs.harvard.edu/abs/2011ApJ...733..111T} {733, 111}

\bibitem[\protect\citeauthoryear{{Valageas}, {Schaeffer}  \& {Silk}}{{Valageas}
  et~al.}{2002}]{Valageas2002}
{Valageas} P.,  {Schaeffer} R.,   {Silk} J.,  2002, \mn@doi [\aap]
  {10.1051/0004-6361:20020548}, \href
  {http://adsabs.harvard.edu/abs/2002A26A...388..741V} {388, 741}

\bibitem[\protect\citeauthoryear{{Verner}, {Verner}  \& {Ferland}}{{Verner}
  et~al.}{1996}]{verner1996}
{Verner} D.~A.,  {Verner} E.~M.,   {Ferland} G.~J.,  1996, \mn@doi [Atomic Data
  and Nuclear Data Tables] {10.1006/adnd.1996.0018}, \href
  {http://adsabs.harvard.edu/abs/1996ADNDT..64....1V} {64, 1}

\bibitem[\protect\citeauthoryear{{Wakker} \& {Savage}}{{Wakker} \&
  {Savage}}{2009}]{Wakker2009}
{Wakker} B.~P.,  {Savage} B.~D.,  2009, \mn@doi [\apjs]
  {10.1088/0067-0049/182/1/378}, \href
  {http://adsabs.harvard.edu/abs/2009ApJS..182..378W} {182, 378}

\bibitem[\protect\citeauthoryear{{Wakker}, {Hernandez}, {French}, {Kim},
  {Oppenheimer}  \& {Savage}}{{Wakker} et~al.}{2015}]{Wakker2015}
{Wakker} B.~P.,  {Hernandez} A.~K.,  {French} D.~M.,  {Kim} T.-S.,
  {Oppenheimer} B.~D.,   {Savage} B.~D.,  2015, \mn@doi [\apj]
  {10.1088/0004-637X/814/1/40}, \href
  {http://adsabs.harvard.edu/abs/2015ApJ...814...40W} {814, 40}

\bibitem[\protect\citeauthoryear{{Werk}, {Prochaska}, {Thom}, {Tumlinson},
  {Tripp}, {O'Meara}  \& {Peeples}}{{Werk} et~al.}{2013}]{Werk2013}
{Werk} J.~K.,  {Prochaska} J.~X.,  {Thom} C.,  {Tumlinson} J.,  {Tripp} T.~M.,
  {O'Meara} J.~M.,   {Peeples} M.~S.,  2013, \mn@doi [\apjs]
  {10.1088/0067-0049/204/2/17}, \href
  {http://adsabs.harvard.edu/abs/2013ApJS..204...17W} {204, 17}

\bibitem[\protect\citeauthoryear{{Werk} et~al.,}{{Werk}
  et~al.}{2016}]{Werk2016}
{Werk} J.~K.,  et~al., 2016, \mn@doi [\apj] {10.3847/1538-4357/833/1/54}, \href
  {http://adsabs.harvard.edu/abs/2016ApJ...833...54W} {833, 54}

\bibitem[\protect\citeauthoryear{{Westra}, {Geller}, {Kurtz}, {Fabricant}  \&
  {Dell'Antonio}}{{Westra} et~al.}{2010}]{Westra2010}
{Westra} E.,  {Geller} M.~J.,  {Kurtz} M.~J.,  {Fabricant} D.~G.,
  {Dell'Antonio} I.,  2010, \mn@doi [\pasp] {10.1086/657452}, \href
  {http://adsabs.harvard.edu/abs/2010PASP..122.1258W} {122, 1258}

\bibitem[\protect\citeauthoryear{{Wright}}{{Wright}}{2006}]{Wright2006}
{Wright} E.~L.,  2006, \mn@doi [\pasp] {10.1086/510102}, \href
  {http://adsabs.harvard.edu/abs/2006PASP..118.1711W} {118, 1711}

\makeatother
\end{thebibliography}
\end{document}